\documentclass[11pt]{article}
\usepackage{amssymb,amsmath,amsfonts}
\usepackage{graphicx}
\usepackage{graphics}
\usepackage{epsfig}
\usepackage{slashed}

 

\begin{document}

\title{Fermion renormalized vertex functions, effective mass, and condensate
in an external Yang-Mills gauge field}
\author{V. V. Parazian\thanks{%
E-mail: vparazian@gmail.com } \\
\textit{Institute of Applied Problems of Physics,}\\
\textit{25 Nersessian Street, 0014 Yerevan, Armenia}}
\maketitle

\begin{abstract}
We investigate the renormalized fermion-gluon vertex, the fermion
effective mass, and the fermion condensate when the fermion
propagates in an external Yang-Mills gauge field. We use an exact
Green's function for the Dirac operator in a non-Abelian plane-wave
gauge field to construct the renormalized vertex function, calculate
the on-shell fermion self-energy, and the background-dependent
condensate. We consider both the background and operator fields in the axial gauge $k^{\mu }%
\mathcal{A}_{\mu }^{a}=0$, thereby preserving the gauge. Its
applications to strong-field QCD and non-Abelian Schwinger physics
are discussed.
\end{abstract}

\section{Introduction}

\label{Introduction}

Changes in propagators, vertices caused by external fields are significant
for coherent color-field models, early heavy-ion physics, and cosmological
scenarios involving large classical gauge fields. A classical solution of
the gauge field with quantized gluon fluctuations is very helpful.

Quantum field theory in specific backgrounds dates back to Schwinger's
external-field formulation \cite{Schwinger1951}, which described how strong
classical fields alter perturbation theory. While the Abelian case has been
extensively developed, particularly in strong-field QED \cite{Ritus1972},
\cite{Narozhny1979}, \cite{Nikishov1985}, extending it to Yang-Mills theory
is more complex and involved because of gauge self-interactions and the
color structure. The background-field method provides a consistent framework
in which Green's functions and effective actions can be constructed while
maintaining gauge invariance with respect to the classical configuration
\cite{Abbott1981}, \cite{Abbott1982}.

In noncovariant gauges, however, perturbation theory introduces
gauge-dependent singularities in the gluon propagator. Their consistent
treatment requires a causal prescription; throughout this work, we employ
the Mandelstam-Leibbrandt regularization \cite{Mandelstam1983}, \cite%
{Leibbrandt1984}. Recent progress shows that it is possible to do exact or
semi-exact analysis in plane-wave and other nontrivial Yang-Mills
backgrounds \cite{Adamo2023}. This allows direct calculation of radiative
quantities in analytical background solutions.

Background fields also alter the vacuum's order parameters. Less attention
has been given to the simultaneous treatment of condensates, self-energies,
and vertex corrections in analytically tractable non-Abelian plane-wave
backgrounds.

In this work, we analyze a fermion propagating in an external plane-wave
Yang-Mills field that satisfies the classical equations of motion \cite%
{Coleman1977}. Based on the exact fermion propagator found in \cite%
{Parazian2025}, we calculate the one-loop fermion-gluon vertex, the
renormalized fermion self-energy, the associated effective mass shift in the
external Yang-Mills gauge field.

A single gauge condition is applied to both the background and the operator
part of the external field by choosing the axial gauge for the external
Yang-Mills gauge field. The Mandelstam-Leibbrandt prescription is used to
handle all singular structures. For the effective mass shift, using the
exact fermionic propagator in the external Yang-Mills plane wave enables us
to separate the contribution from the field-free part from that caused by
the external field. In particular, the effective mass is directly determined
by the renormalized self-energy constructed from the exact propagator, and
background corrections appear as characteristic oscillatory patterns
associated with the plane-wave solution of the Yang-Mills equation.

The paper is organized as follows. Section 2 reviews the gluon propagator in
the chosen gauge. Section 3 discusses the fermion-gluon renormalized vertex
functions (the one-loop vertex correction). Section 4 derives the fermion's
effective mass from the renormalized self-energy. Section 5 analyzes the
fermion condensate. Section 6 presents the discussion and outlook.

Portions of the manuscript were edited with the assistance of an AI-based
language model. The author holds full responsibility for the content,
analysis, and conclusions provided. The scientific content is exclusively
the author's responsibility.

\section{Propagator of the Yang-Mills gauge field}

\label{YangMillsprop}

The external Yang-Mills gauge field $A_{a}^{\mu }$ satisfies the equations%
\begin{equation}
\partial _{\mu }F_{a}^{\nu \mu }\left( x\right) -gf_{ab}{}^{c}A_{\mu
}^{b}F_{c}^{\nu \mu }\left( x\right) =0,  \label{YMeq}
\end{equation}%
\begin{equation}
F_{a}^{\nu \mu }\left( x\right) =\partial ^{\nu }A_{a}^{\mu }-\partial ^{\mu
}A_{a}^{\nu }\left( x\right) -gf_{a}{}^{bc}A_{b}^{\nu }\left( x\right)
A_{c}^{\mu }\left( x\right) ,  \label{Fmunudef}
\end{equation}%
where $x^{\mu }=\left( x^{0},\vec{x}\right) $, $\partial _{\mu }=\left(
\partial /\partial t,\nabla \right) $, Roman letters are used to number the
basis in the space of the $SU(N)$ fundamental representation group; $%
a,b,c=1,\cdots ,N^{2}-1$, diag$\left( G^{\mu \nu }\right) =\left(
1,-1,-1,-1\right) $, $G^{\mu \nu }$ is the metric tensor. For equations (\ref%
{YMeq}) and (\ref{Fmunudef}), the solution can be expressed as a plane wave
moving along the light cone \cite{Coleman1977}.%
\begin{equation}
A_{+}^{a}\left( x\right) =f^{a}\left( x^{+}\right) x^{1}+g^{a}\left(
x^{+}\right) x^{2},\;A_{-}^{a}=A_{1}^{a}=A_{2}^{a}=0,  \label{Acoleman}
\end{equation}%
\begin{equation}
A_{\mu }^{a}\left( x\right) =A_{\mu }^{a}\left( qx\right) ,\;q_{\mu }q^{\mu
}=0,  \label{Amuq}
\end{equation}%
The axial gauge is proper in this case \cite{Coleman1977}.%
\begin{equation}
\partial ^{\mu }A_{\mu }^{a}\left( x\right) =q^{\mu }\dot{A}_{\mu
}^{a}\left( x\right) =\partial ^{+}A_{+}^{a}\left( x\right) =0\;\Rightarrow
\;q^{\mu }A_{\mu }^{a}\left( x\right) =0.  \label{gaugefixing}
\end{equation}%
Let us split the external Yang-Mills gauge field into the background field $%
\mathcal{A}_{\mu }^{a}\left( x\right) $\ and the operator part $\mathsf{A}%
_{\mu }^{a}\left( x\right) $\ as $A_{\mu }^{a}\left( x\right) =\mathcal{A}%
_{\mu }^{a}\left( x\right) +\mathsf{A}_{\mu }^{a}\left( x\right) $. The
operator part describes the quantum fluctuations. In the case of the plane
wave and the axial gauge, we can write the following decomposition: \
\begin{equation}
\mathsf{A}_{a}^{\mu }\left( x\right) =\int \frac{d^{3}q}{\left( 2\pi \right)
^{3}}\frac{1}{\sqrt{2E_{q}}}\sum_{\lambda =1}^{2}\left[ \epsilon _{a}^{\mu
}\left( \vec{q},\lambda \right) \hat{c}_{\lambda a}\left( \vec{q}\right)
\exp \left( -iqx\right) +\epsilon _{a}^{\ast \mu }\left( \vec{q},\lambda
\right) \hat{c}_{\lambda a}^{\dag }\left( \vec{q}\right) \exp \left(
iqx\right) \right] ,  \label{decomposition}
\end{equation}%
where $\hat{c}_{\lambda a}\left( \vec{q}\right) $ and $\hat{c}_{\lambda
a}^{\dag }\left( \vec{q}\right) $ are the operators of cancellation and
creation, respectively. $\epsilon _{a}^{\mu }\left( \vec{q},\lambda \right) $
describes the polarization of the Yang-Mills gauge field. We have the
following conditions for them:%
\begin{align}
\left[ \hat{c}_{\lambda a}^{\dag }\left( \vec{q}\right) ,\hat{c}_{\lambda
^{\prime }b}\left( \vec{q}^{\prime }\right) \right] =&-\delta _{ab}\eta
_{\lambda \lambda ^{\prime }}\left( 2\pi \right) ^{3}\delta ^{\left(
3\right) }\left( \vec{q}-\vec{q}^{\prime }\right) ,  \notag \\
\epsilon _{a}^{\ast \mu }\left( \vec{q},\lambda \right) \epsilon _{\mu
a}\left( \vec{q},\lambda ^{\prime }\right) =&\eta _{\lambda \lambda ^{\prime
}},\;\epsilon _{a}^{\mu }\left( \vec{q},\lambda \right) q_{\mu }=0,  \notag
\\
\langle 0|\hat{c}_{\lambda a}\left( \vec{q}\right) \hat{c}_{\lambda a}^{\dag
}\left( \vec{q}\right) |0\rangle =&1,  \label{opercond}
\end{align}%
In this gauge, we have that only 2 physical polarizations propagate, and%
\begin{equation}
\sum_{\lambda =1}^{2}\epsilon _{\nu }^{a}\left( -\vec{q},\lambda \right)
\epsilon _{\mu }^{\ast b}\left( -\vec{q},\lambda \right) =\delta ^{ab}\left[
-g_{\mu \nu }+\frac{q^{\mu }n^{\nu }+q^{\nu }n^{\mu }}{q\cdot n}+\frac{n^{2}%
}{\left( q\cdot n\right) ^{2}}q^{\mu }q^{\nu }\right] .  \label{polarsum}
\end{equation}%
where $n^{\mu }$ is a fixed unit 4-vector.

The gluon propagator in axial gauges is%
\begin{equation}
D_{\mu \nu }^{ab}\left( x,x^{\prime }\right) =i\left( \theta \left(
x^{0}-x^{0\prime }\right) \langle 0|\mathsf{A}_{\mu }^{a}\left( x\right)
\mathsf{A}_{\nu }^{b}\left( x^{\prime }\right) |0\rangle +i\theta \left(
x^{0\prime }-x^{0}\right) \langle 0|\mathsf{A}_{\nu }^{a}\left( x^{\prime
}\right) \mathsf{A}_{\mu }^{b}\left( x\right) |0\rangle \right) .
\label{gluonprop}
\end{equation}%
Substituting (\ref{decomposition}) and (\ref{polarsum}) into the (\ref%
{gluonprop}) and utilizing (\ref{opercond}), we derive the following
expression for the gluon propagator (see, for instance, \cite{Leibbrandt1987}%
, \cite{Joglekar2000})%
\begin{equation}
D_{\mu \nu }^{ab}\left( x,x^{\prime }\right) =-\int_{C}\frac{d^{4}q}{\left(
2\pi \right) ^{4}}\frac{e^{-q\left( x-x^{\prime }\right) }}{q^{2}+i\epsilon }%
\delta ^{ab}\left[ -g_{\mu \nu }+\frac{q^{\mu }n^{\nu }+q^{\nu }n^{\mu }}{%
q\cdot n}+\frac{n^{2}}{\left( q\cdot n\right) ^{2}}q^{\mu }q^{\nu }\right] ,
\label{glprop1}
\end{equation}%
where the contour $C$ is the standard contour of integration. Through this
decision, gauge redundancy is removed without the typical introduction of
Fadeev-Popov ghosts (see, for instance, \cite{Greiner2007}). Using the
additional vector light-like $n^{2}=0$, we get the expression:%
\begin{equation}
\tilde{D}_{\mu \nu }^{ab}\left( x,x^{\prime }\right) =-iD_{\mu \nu
}^{ab}\left( x,x^{\prime }\right) =\int_{C}\frac{d^{4}q}{\left( 2\pi \right)
^{4}}\frac{i\delta ^{ab}}{q^{2}+i\epsilon }\left[ -g_{\mu \nu }+\frac{q^{\mu
}n^{\nu }+q^{\nu }n^{\mu }}{q\cdot n}\right] e^{-q\left( x-x^{\prime
}\right) }.  \label{glprop2}
\end{equation}

According to a prescription independently suggested by Mandelstam \cite%
{Mandelstam1983} and Leibbrandt \cite{Leibbrandt1984} (ML)%
\begin{equation}
\frac{1}{q\cdot n}\rightarrow \frac{1}{\left[ qn\right] }=\lim_{\epsilon
_{3}\rightarrow 0}\frac{n^{\ast }q}{\left( n^{\ast }q\right) \left( q\cdot
n\right) +i\epsilon _{3}}  \label{ML}
\end{equation}%
where $n^{\ast }$ is the vector conjugate to the vector $n$, we have the
following expression for $D_{\mu \nu }^{ab}\left( x,x^{\prime }\right) $ and
$\tilde{D}_{\mu \nu }^{ab}\left( x,x^{\prime }\right) $
\begin{equation}
\tilde{D}_{\mu \nu }^{ab}\left( x,x^{\prime }\right) =-iD_{\mu \nu
}^{ab}\left( x,x^{\prime }\right) =\int_{C}\frac{d^{4}q}{\left( 2\pi \right)
^{4}}\frac{i\delta ^{ab}}{q^{2}+i\epsilon }\left[ -g_{\mu \nu }+\frac{\left(
q^{\mu }n^{\nu }+q^{\nu }n^{\mu }\right) n^{\ast }q}{\left( n^{\ast
}q\right) \left( q\cdot n\right) +i\epsilon _{3}}\right] e^{-q\left(
x-x^{\prime }\right) }.  \label{gluonpropML}
\end{equation}

\section{Renormalized vertex functions}

\label{vertexfunctionsselfenergy}

The one-loop vertex correction in momentum space is%
\begin{align}
v^{\dagger }\bar{u}\left( p^{\prime }\right) \Gamma ^{c\mu }\left(
p,k\right) u\left( p\right) v=& v^{\dagger }\bar{u}\left( p^{\prime }\right)
\left( -ig\gamma ^{\alpha }T^{a}\right) \tilde{G}_{F}\left( q+k\right)
\gamma ^{\mu }T^{c}\tilde{G}_{F}\left( q\right)  \notag \\
& \times \left( -ig\gamma ^{\beta }T^{b}\right) u\left( p\right) v\tilde{D}%
_{\alpha \beta }^{ab}\left( p-q\right) ,  \label{vertexmomspace}
\end{align}%
where $p^{\prime }=p+k$, $u\left( p\right) $ and $v$ are spinors (indices
omitted), which are elements of the spaces of the appropriate
representations \cite{Koshelkin2010}, and for $u_{\sigma }\left( p\right) $,
we use the following normalization:
\begin{equation}
\bar{u}_{\sigma }\left( p\right) u_{\lambda }\left( p^{\prime }\right) =\pm
2m\delta _{\sigma \lambda }\delta _{pp^{\prime }},\ p^{2}=m^{2},
\label{fermnorm}
\end{equation}%
where the Dirac scalar production of the spinors $u_{\sigma }\left( p\right)
$ and $u_{\sigma }\left( -p\right) $, respectively, is represented by the
plus and minus signs. The spinor $v_{\alpha }$ is normalized by condition $%
v_{\alpha }^{\dag }v_{\beta }=\delta _{\alpha \beta }$.

For $\tilde{G}_{F}\left( x,y\right) $, we used results from \cite%
{Parazian2025}.
\begin{equation}
\tilde{G}_{F}\left( p\right) =\frac{i\left( \slashed{p}+m\right) U\left(
p\right) }{p^{2}-m^{2}+i\epsilon },  \label{Fermionpropmomspace}
\end{equation}%
\begin{eqnarray}
U\left( p\right) &=&U\left( p,\varphi ,\varphi ^{\prime }\right) =\cos
\left( \theta \left( p,\varphi \right) \right) \cos \left( \theta \left(
p,\varphi ^{\prime }\right) \right)  \notag \\
&&\times \left\{ 1+\frac{\tan \left( \theta \left( p,\varphi ^{\prime
}\right) \right) }{\theta \left( p,\varphi ^{\prime }\right) }\frac{g\left(
\left( \gamma ^{\sigma }\right) ^{\dagger }\mathcal{A}_{\sigma }^{e}\left(
\varphi ^{\prime }\right) \right) \left( \left( \gamma ^{\rho }\right)
^{\dagger }k_{\rho }\right) }{2\left( pk\right) }T_{e}\right.  \notag \\
&&\left. +\frac{g\left( \gamma ^{\nu }k_{\nu }\right) \left( \gamma
^{\lambda }\mathcal{A}_{\lambda }^{b}\left( \varphi \right) \right) }{%
2\left( pk\right) }\frac{\tan \left( \theta \left( p,\varphi \right) \right)
}{\theta \left( p,\varphi \right) }T_{b}+\frac{g\left( \gamma ^{\nu }k_{\nu
}\right) \left( \gamma ^{\lambda }\mathcal{A}_{\lambda }^{b}\left( \varphi
\right) \right) }{2\left( pk\right) }\frac{\tan \left( \theta \left(
p,\varphi \right) \right) }{\theta \left( p,\varphi \right) }\right.  \notag
\\
&&\left. \times \frac{\tan \left( \theta \left( p,\varphi ^{\prime }\right)
\right) }{\theta \left( p,\varphi ^{\prime }\right) }\frac{g\left( \left(
\gamma ^{\sigma }\right) ^{\dagger }\mathcal{A}_{\sigma }^{e}\left( \varphi
^{\prime }\right) \right) \left( \left( \gamma ^{\rho }\right) ^{\dagger
}k_{\rho }\right) }{2\left( pk\right) }T_{b}T_{e}\right\} ,  \label{Udef}
\end{eqnarray}%
\begin{equation}
\theta \left( p\right) =\theta \left( p,\varphi \right) =\frac{g}{\left(
pk\right) }\sqrt{\frac{1}{2N}}\left( \int_{0}^{\varphi }d\varphi ^{\prime
\prime }\left( \mathcal{A}_{\mu }^{a}\left( \varphi ^{\prime \prime }\right)
p^{\mu }\right) \int_{0}^{\varphi }d\varphi ^{\prime \prime \prime }\left(
\mathcal{A}_{a}^{\mu }\left( \varphi ^{\prime \prime \prime }\right) p_{\mu
}\right) \right) ^{\frac{1}{2}},  \label{thetadef}
\end{equation}

$\varphi =kx,\ k^{\mu }\mathcal{A}_{\mu }^{a}\left( x\right) =0$. According
to (\ref{gluonpropML}), the gluon propagator given in momentum space is%
\begin{equation}
\tilde{D}_{\mu \nu }^{ab}\left( q\right) =\frac{i\delta ^{ab}}{%
q^{2}+i\epsilon _{3}}\left[ -g_{\alpha \beta }+\frac{\left( q_{\alpha
}n_{\beta }+q_{\beta }n_{\alpha }\right) \left( n^{\ast }\cdot q\right) }{%
\left( n^{\ast }\cdot q\right) \left( q\cdot n\right) +i\varepsilon }\right]
.  \label{gluonmomspace}
\end{equation}%
Substituting (\ref{Fermionpropmomspace}) and (\ref{gluonpropML}) into (\ref%
{vertexmomspace}), we obtain%
\begin{eqnarray}
\Gamma ^{c\mu }\left( p,k\right) &=&\int \frac{d^{4}r}{\left( 2\pi \right)
^{4}}v^{\dagger }\bar{u}\left( p^{\prime }\right) \left( -ig\gamma ^{\alpha
}T^{a}\right) \frac{i\left( \slashed{P}+m\right) U\left( P\right) }{%
P^{2}-m^{2}+i\varepsilon }\gamma ^{\mu }T^{c}\frac{i\left( \slashed{r}%
+m\right) U\left( r\right) }{r^{2}-m^{2}+i\varepsilon }  \notag \\
&&\times \left( -ig\gamma ^{\beta }T^{b}\right) u\left( p\right) v\frac{%
i\delta ^{ab}}{q^{2}+i\varepsilon }\left( -g_{\alpha \beta }+\frac{\left(
q_{\alpha }n_{\beta }+q_{\beta }n_{\alpha }\right) q\cdot n^{\ast }}{q\cdot
n^{\ast }q\cdot n+i\varepsilon }\right) .  \label{vertexmomentspaceexact}
\end{eqnarray}%
where $p^{\prime }=p+k$ and loop momentum $r$, and we have denoted $P\equiv
r+k$ and gluon momentum $q\equiv p-r$.

It is clear that as the limit $\mathcal{A\rightarrow }0$, $U\rightarrow 1$,
and we substitute $T^{a}=1$, $g=e$, and $v=1$, which comes from formula (\ref%
{Fermionpropmomspace}),\ we obtain%
\begin{eqnarray}
\Gamma _{El}^{\mu }\left( p,k\right) &=&e^{2}\bar{u}\left( p^{\prime
}\right) \int \frac{d^{4}r}{\left( 2\pi \right) ^{4}}\gamma ^{\alpha }\frac{%
\left( \slashed{P}+m\right) }{P^{2}-m^{2}+i\varepsilon }\gamma ^{\mu }\frac{%
\left( \slashed{r}+m\right) }{r^{2}-m^{2}+i\epsilon }\gamma ^{\beta }  \notag
\\
&&\times \frac{1}{q^{2}+i\varepsilon }\left( -g_{\alpha \beta }+\frac{\left(
q_{\alpha }n_{\beta }+q_{\beta }n_{\alpha }\right) \left( q\cdot n^{\ast
}\right) }{q\cdot n^{\ast }q\cdot n+i\epsilon }\right) u\left( p\right) ,
\label{vertexEl}
\end{eqnarray}%
This matches the one-loop vertex correction in momentum space of
electromagnetic interaction.

We can write the weak-field expansion as $U=1+\mathcal{O}\left( g\mathcal{A}%
\right) +\mathcal{O}\left( \left( g\mathcal{A}\right) ^{2}\right) $, and
substitute it into the expression for the one-loop vertex correction in
momentum space. Thus,%
\begin{eqnarray}
U\left( p,\varphi ,\varphi ^{\prime }\right) &=&\cos \left( \theta \left(
p,\varphi \right) \right) \cos \left( \theta \left( p,\varphi ^{\prime
}\right) \right) \left\{ 1+\left( \cdots \right) _{\mathcal{O}\left( g%
\mathcal{A}\right) }+\left( \cdots \right) _{\mathcal{O}\left( \left( g%
\mathcal{A}\right) ^{2}\right) }\right\} ,  \notag \\
\cos \theta &=&1-\frac{\theta ^{2}}{2}+\mathcal{O}\left( \theta ^{4}\right)
,\;\frac{\tan \theta }{\theta }=1+\frac{\theta ^{2}}{3}+\mathcal{O}\left(
\theta ^{4}\right) ,  \label{expanUtet}
\end{eqnarray}%
thus, we can set $\left( \frac{\tan \theta }{\theta }\right) \rightarrow 1$
at the linear order. We denote%
\begin{eqnarray}
X\left( l;\varphi \right) &=&\frac{g}{2\left( lk\right) }\left( \gamma ^{\nu
}k_{\nu }\right) \left( \gamma ^{\lambda }\mathcal{A}_{\lambda }^{b}\left(
\varphi \right) \right) T_{b},  \notag \\
Y\left( l;\varphi ^{\prime }\right) &=&\frac{g}{2\left( lk\right) }\left(
\left( \gamma ^{\sigma }\right) ^{\dagger }\mathcal{A}_{\sigma }^{e}\left(
\varphi ^{\prime }\right) \right) \left( \left( \gamma ^{\rho }\right)
^{\dagger }k_{\rho }\right) T_{e},  \notag \\
XY\left( l;\varphi ,\varphi ^{\prime }\right) &=&\left( \frac{g}{2\left(
lk\right) }\right) ^{2}\left( \gamma ^{\nu }k_{\nu }\right) \left( \gamma
^{\lambda }\mathcal{A}_{\lambda }^{b}\left( \varphi \right) \right) \left(
\left( \gamma ^{\sigma }\right) ^{\dagger }\mathcal{A}_{\sigma }^{e}\left(
\varphi ^{\prime }\right) \right) \left( \left( \gamma ^{\rho }\right)
^{\dagger }k_{\rho }\right) T_{b}T_{e},  \label{XYdef}
\end{eqnarray}%
so that $X$, $Y=O\left( g\mathcal{A}\right) $. The scalar prefactor
correlation from $\cos \left( \theta \left( p,\varphi \right) \right) \cos
\left( \theta \left( p,\varphi ^{\prime }\right) \right) $ at $\mathcal{O}%
\left( \left( g\mathcal{A}\right) ^{2}\right) $ is color-singlet%
\begin{equation}
Z\left( l;\varphi ,\varphi ^{\prime }\right) \equiv -\frac{1}{2}\left(
\theta \left( l,\varphi \right) ^{2}+\theta \left( l,\varphi ^{\prime
}\right) ^{2}\right) \mathbf{1.}  \label{Zldefinition}
\end{equation}

Then, up to quadratic order, a compact and correct expansion is
\begin{equation}
U\left( l;\varphi ,\varphi ^{\prime }\right) =\mathbf{1}+X\left( l;\varphi
\right) +Y\left( l;\varphi ^{\prime }\right) +X\left( l;\varphi \right)
Y\left( l;\varphi ^{\prime }\right) +Z\left( l;\varphi ,\varphi ^{\prime
}\right) +O\left( \left( g\mathcal{A}\right) ^{3}\right) .  \label{expanU}
\end{equation}%
For $\Gamma ^{c\mu }$, we write $\Gamma ^{c\mu }=\Gamma _{\left( 0\right)
}^{c\mu }+\Gamma _{\left( 1\right) }^{c\mu }+\Gamma _{\left( 2\right)
}^{c\mu }+O\left( \left( g\mathcal{A}\right) ^{3}\right) $, where the
subscripts denote the order in the background amplitude.

For the part corresponding to the zero external field ($U\rightarrow 1$), we
have%
\begin{eqnarray}
\Gamma _{\left( 0\right) }^{c\mu } &=&\left( -ig^{2}\right) \int \frac{d^{4}r%
}{\left( 2\pi \right) ^{4}}v^{\dagger }\bar{u}\left( p^{\prime }\right)
\gamma ^{\alpha }T^{a}\frac{i\left( \slashed{P}+m\right) }{%
P^{2}-m^{2}+i\epsilon }\gamma ^{\mu }T^{c}\frac{i\left( \slashed{r}+m\right)
}{r^{2}-m^{2}+i\epsilon }  \notag \\
&&\times \gamma ^{\beta }T^{a}u\left( p\right) v\frac{i}{q^{2}+i\epsilon }%
H_{\alpha \beta }\left( q\right) ,  \label{Gammacmyuzro}
\end{eqnarray}%
where
\begin{equation}
H_{\alpha \beta }\left( q\right) =\left( -g_{\alpha \beta }+\frac{\left(
q_{\alpha }n_{\beta }+q_{\beta }n_{\alpha }\right) q\cdot n^{\ast }}{q\cdot
n^{\ast }q\cdot n+i\epsilon }\right) .  \label{Halefbetta}
\end{equation}%
Only one of the two $U$'s is expanded to first order for the linear
background component $O\left( g\mathcal{A}\right) $, so we have:%
\begin{eqnarray}
\Gamma _{\left( 1\right) }^{c\mu } &=&\left( -ig^{2}\right) \int \frac{d^{4}r%
}{\left( 2\pi \right) ^{4}}v^{\dagger }\bar{u}\left( p^{\prime }\right)
\gamma ^{\alpha }T^{a}\frac{i\left( \slashed{P}+m\right) }{%
P^{2}-m^{2}+i\epsilon }  \notag \\
&&\left\{ X\left( P,\varphi \right) \gamma ^{\mu }T^{c}\frac{i\left( %
\slashed{r}+m\right) }{r^{2}-m^{2}+i\varepsilon }+Y\left( P,\varphi ^{\prime
}\right) \gamma ^{\mu }T^{c}\frac{i\left( \slashed{r}+m\right) }{%
r^{2}-m^{2}+i\epsilon }\right.  \notag \\
&&\left. +\gamma ^{\mu }T^{c}\frac{i\left( \slashed{r}+m\right) }{%
r^{2}-m^{2}+i\varepsilon }X\left( r,\varphi \right) +\gamma ^{\mu }T^{c}%
\frac{i\left( \slashed{r}+m\right) }{r^{2}-m^{2}+i\epsilon }Y\left(
r,\varphi ^{\prime }\right) \right\}  \notag \\
&&\times \gamma ^{\beta }T^{b}u\left( p\right) v\frac{i\delta ^{ab}}{%
q^{2}+i\epsilon }H_{\alpha \beta }\left( q\right) .  \label{Gammacmu1}
\end{eqnarray}

The quadratic term $\mathcal{O}\left( \left( g\mathcal{A}\right) ^{2}\right)
$ has three structural sources: a second-order insertion on $P$ only: $\left[
XY+Z\right] \left( P\right) $, a second-order insertion on r only: $\left[
XY+Z\right] \left( r\right) $, and a first-order insertion on each line: $%
\left[ X+Y\right] \left( P\right) \left[ X+Y\right] \left( r\right) $.
Second-order terms are%
\begin{eqnarray}
\Gamma _{\left( 2\right) ,P;XY}^{c\mu } &=&\left( -ig^{2}\right) \int \frac{%
d^{4}r}{\left( 2\pi \right) ^{4}}v^{\dagger }\bar{u}\left( p^{\prime
}\right) \gamma ^{\alpha }T^{a}\frac{i\left( \slashed{P}+m\right) }{%
P^{2}-m^{2}+i\epsilon }\left[ XY\left( P;\varphi ,\varphi ^{\prime }\right) %
\right]  \notag \\
&&\times \gamma ^{\mu }T^{c}\frac{i\left( \slashed{r}+m\right) }{%
r^{2}-m^{2}+i\varepsilon }\gamma ^{\beta }T^{b}u\left( p\right) v\frac{%
i\delta ^{ab}}{q^{2}+i\epsilon }H_{\alpha \beta }\left( q\right) ,
\label{Gam2PXY}
\end{eqnarray}%
\begin{eqnarray}
\Gamma _{\left( 2\right) ,P;Z}^{c\mu } &=&\left( -ig^{2}\right) \int \frac{%
d^{4}r}{\left( 2\pi \right) ^{4}}v^{\dagger }\bar{u}\left( p^{\prime
}\right) \gamma ^{\alpha }T^{a}\frac{i\left( \slashed{P}+m\right) }{%
P^{2}-m^{2}+i\epsilon }\left[ Z\left( P;\varphi ,\varphi ^{\prime }\right) %
\right]  \notag \\
&&\times \gamma ^{\mu }T^{c}\frac{i\left( \slashed{r}+m\right) }{%
r^{2}-m^{2}+i\varepsilon }\gamma ^{\beta }T^{b}u\left( p\right) v\frac{%
i\delta ^{ab}}{q^{2}+i\epsilon }H_{\alpha \beta }\left( q\right) ,
\label{Gam2PZ}
\end{eqnarray}%
\begin{eqnarray}
\Gamma _{\left( 2\right) ,r;XY}^{c\mu } &=&\left( -ig^{2}\right) \int \frac{%
d^{4}r}{\left( 2\pi \right) ^{4}}v^{\dagger }\bar{u}\left( p^{\prime
}\right) \gamma ^{\alpha }T^{a}\frac{i\left( \slashed{P}+m\right) }{%
P^{2}-m^{2}+i\epsilon }\gamma ^{\mu }T^{c}  \notag \\
&&\times \frac{i\left( \slashed{r}+m\right) }{r^{2}-m^{2}+i\varepsilon }%
\left[ XY\left( r;\varphi ,\varphi ^{\prime }\right) \right] \gamma ^{\beta
}T^{b}u\left( p\right) v\frac{i\delta ^{ab}}{q^{2}+i\epsilon }H_{\alpha
\beta }\left( q\right) ,  \label{Gam2rXY}
\end{eqnarray}%
\begin{eqnarray}
\Gamma _{\left( 2\right) ,r;Z}^{c\mu } &=&\left( -ig^{2}\right) \int \frac{%
d^{4}r}{\left( 2\pi \right) ^{4}}v^{\dagger }\bar{u}\left( p^{\prime
}\right) \gamma ^{\alpha }T^{a}\frac{i\left( \slashed{P}+m\right) }{%
P^{2}-m^{2}+i\epsilon }\gamma ^{\mu }T^{c}  \notag \\
&&\times \frac{i\left( \slashed{r}+m\right) }{r^{2}-m^{2}+i\varepsilon }%
\left[ Z\left( P;\varphi ,\varphi ^{\prime }\right) \right] \gamma ^{\beta
}T^{b}u\left( p\right) v\frac{i\delta ^{ab}}{q^{2}+i\epsilon }H_{\alpha
\beta }\left( q\right) ,  \label{Gam2rZ}
\end{eqnarray}%
Now we expand $\left( X+Y\right) \left( P\right) \left( X+Y\right) \left(
r\right) $, resulting in four cross-terms. For example:%
\begin{eqnarray}
\Gamma _{\left( 2\right) ,P,X;r,X}^{c\mu } &=&\left( -ig^{2}\right) \int
\frac{d^{4}r}{\left( 2\pi \right) ^{4}}v^{\dagger }\bar{u}\left( p^{\prime
}\right) \gamma ^{\alpha }T^{a}\frac{i\left( \slashed{P}+m\right) }{%
P^{2}-m^{2}+i\epsilon }X\left( P,\varphi \right)  \notag \\
&&\times \gamma ^{\mu }T^{c}\frac{i\left( \slashed{r}+m\right) }{%
r^{2}-m^{2}+i\varepsilon }X\left( r,\varphi \right) \gamma ^{\beta
}T^{b}u\left( p\right) v\frac{i\delta ^{ab}}{q^{2}+i\epsilon }H_{\alpha
\beta }\left( q\right)  \label{G2X+YPX+Yr}
\end{eqnarray}%
For $\Gamma _{\left( 2\right) ,P,X;r,Y}^{c\mu }$, $\Gamma _{\left( 2\right)
,P,Y;r,X}^{c\mu }$, $\Gamma _{\left( 2\right) ,P,Y;r,Y}^{c\mu }$, we can
write similarly.

Now, we analyze the periodic plane-wave background: $\mathcal{A}_{\mu
}^{a}\left( \varphi \right) =\varepsilon _{\mu }^{a}\cos \left( \varphi
\right) $, $\varphi =\kappa \cdot x$, $\kappa ^{2}=0$, $\kappa \cdot
\varepsilon ^{a}=0$. To prevent collisions, we use $k$ to represent the
emitted gluon momentum and $\kappa $ to denote the background wavevector.
Every background-dependent object built from $\mathcal{A}\left( \varphi
\right) $ admits a Fourier series in $\varphi $ (and $\varphi ^{\prime }$)
due to the periodicity of $\mathcal{A}\left( \varphi \right) $. In
particular, the standard Floquet form can be used to describe the exact
dressed fermion factors for a monochromatic plane wave%
\begin{equation}
U\left( l;\varphi ,\varphi ^{\prime }\right) =\sum_{n=-\infty }^{\infty
}U_{n}\left( l\right) e^{-in\left( \varphi -\varphi ^{\prime }\right) },
\label{UexpandFloquet}
\end{equation}%
with coefficients $U_{n}\left( l\right) $ expressed in terms of Bessel
functions $J_{n}$ of an \textquotedblleft intensity
parameter\textquotedblright\ (the non-Abelian generalization of the Volkov
parameter). At weak fields, this reduces to a finite number of harmonics at
each order: $O\left( \varepsilon \right) $: only $n=\pm 1$, $O\left(
\varepsilon ^{2}\right) $: only $n=0,\pm 2$, etc. When such a Fourier
expansion is substituted into the coordinate-space amplitude, and a Fourier
transform to momentum space is performed, the phase factor $e^{-i\kappa
\cdot x}$ results in shifted momentum-conserving delta functions. We can
write,%
\begin{equation}
\int d^{4}xe^{i\left( p^{\prime }-p-k\right) \cdot x}e^{-in\kappa \cdot
x}=\left( 2\pi \right) ^{4}\delta ^{\left( 4\right) }\left( p^{\prime
}-p-k-n\kappa \right) .  \label{delta4exter}
\end{equation}%
Therefore, the precise selection rule is $\left( 2\pi \right) ^{4}\delta
^{\left( 4\right) }\left( p^{\prime }-p-k-n\kappa \right) $ for each
harmonic $n$. Momentum conservation is approximately $p^{\prime }\approx
p+k+n\kappa $.

We denote%
\begin{equation}
\alpha \left( l\right) \equiv \frac{g}{l\cdot \kappa }\sqrt{\frac{\left(
l\cdot \varepsilon ^{a}\right) \left( l\cdot \varepsilon _{a}\right) }{2N}}.
\label{alfal}
\end{equation}%
We are using the two harmonic weights: $f_{s}\left( \alpha \left( l\right)
\right) $= even-harmonic weight from $\cos \left( \alpha _{l}\sin \left(
\varphi \right) \right) $, $c_{s}\left( \alpha _{l}\right) $= odd-harmonic
weight from $\cos \left( \varphi \right) \cos \left( \alpha \left( l\right)
\sin \left( \varphi \right) \right) $. We define the selectors $\mathcal{E}%
_{s}=\frac{1+\left( -1\right) ^{s}}{2}$ (1 if $s$ even), $\mathcal{O}_{s}=%
\frac{1-\left( -1\right) ^{s}}{2}$ (1 if $s$ odd). We have the Fourier
expansion for $\cos \left( \alpha \left( l\right) \sin \left( \varphi
\right) \right) $: $\cos \left( \alpha \left( l\right) \sin \left( \varphi
\right) \right) =\sum_{s\in Z}f_{s}\left( \alpha \left( l\right) \right)
e^{is\varphi }$, with explicit parity selection $f_{s}\left( \alpha \left(
l\right) \right) =\mathcal{E}_{s}J_{s}\left( \alpha \left( l\right) \right) $%
, where $J_{s}$ is the Bessel function. So $f_{s}=0$ for odd $s$.

By definition, $\cos \left( \varphi \right) \cos \left( \alpha \left(
l\right) \sin \left( \varphi \right) \right) =\sum_{s\in Z}c_{s}\left(
\alpha \left( l\right) \right) e^{is\varphi }$, and we get the the shift
identity%
\begin{equation}
c_{s}\left( \alpha \left( l\right) \right) =\frac{1}{2}\left( f_{s-1}\left(
\alpha \left( l\right) \right) +f_{s+1}\left( \alpha \left( l\right) \right)
\right) =\frac{1}{2}\left( \mathcal{E}_{s-1}J_{s-1}\left( \alpha \left(
l\right) \right) +\mathcal{E}_{s+1}J_{s+1}\left( \alpha \left( l\right)
\right) \right) .  \label{shiftident}
\end{equation}%
So $c_{s}=0$ for even $s$. In the weak-field (small $\alpha \left( l\right)
\sim g\varepsilon $) version, using the expansion of the Bessel function,
for even weights, we get%
\begin{equation}
f_{0}\left( \alpha \right) =1-\frac{\alpha ^{2}}{4}+\mathcal{O}\left( \alpha
^{4}\right) ,\;f_{\pm 2}=\frac{\alpha ^{2}}{8}+\mathcal{O}\left( \alpha
^{4}\right) ,\;f_{|s|\geqslant 4}=\mathcal{O}\left( \alpha ^{|s|}\right) ,
\label{evenveights}
\end{equation}%
and $f_{odd}=0$. For odd weights, $c_{s}$, we get%
\begin{eqnarray}
c_{\pm 1}\left( \alpha \right) &=&\frac{1}{2}\left( f_{0}+f_{\pm 2}\right) =%
\frac{1}{2}\left( 1-\frac{\alpha ^{2}}{4}+\frac{\alpha ^{2}}{8}\right) +%
\mathcal{O}\left( \alpha ^{4}\right) =\frac{1}{2}-\frac{\alpha ^{2}}{16}+%
\mathcal{O}\left( \alpha ^{4}\right) ,  \notag \\
c_{\pm 3}\left( \alpha \right) &=&\frac{1}{2}\left( f_{\pm 2}+f_{\pm
4}\right) =\frac{\alpha ^{2}}{16}+\mathcal{O}\left( \alpha ^{4}\right) ,
\label{oddweights}
\end{eqnarray}%
and $c_{even}=0$. So, to $O\left( \alpha ^{2}\right) $, only $c_{\pm
1}\left( \alpha \right) \approx \frac{1}{2}-\frac{\alpha ^{2}}{16}$, $c_{\pm
3}\left( \alpha \right) \approx \frac{\alpha ^{2}}{16}$ matter.

Now, we define the linear operator insertion (Dirac+color)%
\begin{equation}
\mathcal{O}^{\left( 1\right) }\left( l\right) \equiv \frac{g}{2\left( l\cdot
\kappa \right) }\left( \left( \gamma ^{\nu }\kappa _{\nu }\right) \left(
\gamma ^{\lambda }\mathcal{\varepsilon }_{\lambda }^{b}\right) T_{b}+\left(
\left( \gamma ^{\sigma }\right) ^{\dagger }\mathcal{\varepsilon }_{\sigma
}^{e}\right) \left( \left( \gamma ^{\rho }\right) ^{\dagger }\kappa _{\rho
}\right) T_{e}\right) .  \label{O1ldef}
\end{equation}%
Then the weak-field Bessel-dressed linear harmonic is%
\begin{eqnarray}
&&\Gamma _{\left( 1\right) ,s}^{c\mu }\left( p,k\right) =\left(
-ig^{2}\right) \int \frac{d^{4}r}{\left( 2\pi \right) ^{4}}\frac{iH_{\alpha
\beta }\left( q\right) }{D_{P}D_{r}D_{g}}  \notag \\
&&\times \left[ c_{s}\left( \alpha \left( _{P}\right) \right) \gamma
^{\alpha }T^{a}\left( i\left( \slashed{P}+m\right) \right) \mathcal{O}%
^{\left( 1\right) }\left( P\right) \gamma ^{\mu }T^{c}\left( i\left( %
\slashed{r}+m\right) \right) \gamma ^{\beta }T^{a}\right.  \notag \\
&&\left. +c_{s}\left( \alpha \left( r\right) \right) \gamma ^{\alpha
}T^{a}\left( i\left( \slashed{P}+m\right) \right) \gamma ^{\mu }T^{c}\left(
i\left( \slashed{r}+m\right) \right) \mathcal{O}^{\left( 1\right) }\left(
r\right) \gamma ^{\beta }T^{a}\right] ,  \label{Gamma1s}
\end{eqnarray}%
where $q=p-r$,$\;P=r+k$,$\;D_{P}=P^{2}-m^{2}+i\epsilon $, $%
D_{r}=r^{2}-m^{2}+i\epsilon $,$\;D_{g}=q^{2}+i\epsilon $, with parity
selection $c_{s}=0$ for even $s$.

At $\mathcal{O}\left( \left( g\varepsilon \right) ^{2}\right) $, there are
three contributions. We examine the $XY$ insertion (even harmonics, uses $%
f_{s}$). We define%
\begin{equation}
\mathcal{O}^{\left( 2\right) }\left( l\right) \equiv \left( \frac{g}{2\left(
l\cdot \kappa \right) }\right) ^{2}\left( \gamma ^{\nu }\kappa _{\nu
}\right) \left( \gamma ^{\lambda }\mathcal{\varepsilon }_{\lambda
}^{b}\right) \left( \left( \gamma ^{\sigma }\right) ^{\dagger }\mathcal{%
\varepsilon }_{\sigma }^{e}\right) \left( \left( \gamma ^{\rho }\right)
^{\dagger }\kappa _{\rho }\right) T_{b}T_{e}  \label{O2ldef}
\end{equation}%
Then%
\begin{eqnarray}
&&\Gamma _{\left( 2\right) ,s}^{c\mu }|_{\mathcal{O}^{\left( 2\right)
}}=\left( -ig^{2}\right) \int \frac{d^{4}r}{\left( 2\pi \right) ^{4}}\frac{%
iH_{\alpha \beta }\left( q\right) }{D_{P}D_{r}D_{g}}  \notag \\
&&\left[ f_{s}\left( \alpha \left( P\right) \right) \gamma ^{\alpha
}T^{a}\left( i\left( \slashed{P}+m\right) \right) \mathcal{O}^{\left(
2\right) }\left( P\right) \gamma ^{\mu }T^{c}\left( i\left( \slashed{r}%
+m\right) \right) \gamma ^{\beta }T^{a}\right.  \notag \\
&&\left. +f_{s}\left( \alpha \left( r\right) \right) \gamma ^{\alpha
}T^{a}\left( i\left( \slashed{P}+m\right) \right) \gamma ^{\mu }T^{c}\left(
i\left( \slashed{r}+m\right) \right) \mathcal{O}_{XY}^{\left( 2\right)
}\left( r\right) \gamma ^{\beta }T^{a}\right] ,  \label{Gamma2ssame}
\end{eqnarray}%
with parity selection $f_{s}\left( \alpha \right) =0$ for odd $s$.

Now, let's analyze the case where one inserts one linear term on each line
(even harmonics, convolution of odd weights). This has weight $%
c_{s_{1}}\left( \alpha _{P}\right) c_{s_{2}}\left( \alpha _{r}\right) $ with
$s_{1}+s_{2}=s$. Explicitly:%
\begin{eqnarray}
&&\Gamma _{\left( 2\right) ,s}^{c\mu }|_{\mathcal{O}^{\left( 1\right) }%
\mathcal{O}^{\left( 1\right) }}=\left( -ig^{2}\right) \int \frac{d^{4}r}{%
\left( 2\pi \right) ^{4}}\frac{iH_{\alpha \beta }\left( q\right) }{%
D_{P}D_{r}D_{g}}  \notag \\
&&\times \sum_{s_{1}+s_{2}=s}c_{s_{1}}\left( \alpha \left( P\right) \right)
c_{s2}\left( \alpha r\right) \gamma ^{\alpha }T^{a}\left( i\left( \slashed{P}%
+m\right) \right)  \notag \\
&&\times \mathcal{O}^{\left( 1\right) }\left( P\right) \gamma ^{\mu
}T^{c}\left( i\left( \slashed{r}+m\right) \right) \mathcal{O}^{\left(
1\right) }\left( r\right) \gamma ^{\beta }T^{a},  \label{Gamma2scross}
\end{eqnarray}%
and since $c$ is odd-only, the sum automatically forces $s$ to be even: $%
\Gamma _{\left( 2\right) ,s}^{c\mu }|_{\mathcal{O}^{\left( 1\right) }%
\mathcal{O}^{\left( 1\right) }}=0$ for odd $s$. At order $O\left( \alpha
^{2}\right) $, only $s_{1}=\pm 1$ and $s_{2}=\mp 1$ contribute to $s=0$,
while $s_{1}=\pm 1$ and $s_{2}=\pm 1$ contribute to $s=\pm 2$, with $c_{\pm
1}\approx \frac{1}{2}$. For the monochromatic background $\mathcal{A}_{\mu
}^{a}\left( \varphi \right) =\varepsilon _{\mu }^{a}\cos \left( \varphi
\right) $, we have, for any momentum $\ell $, $\theta \left( l,\varphi
\right) =\alpha _{l}\sin \left( \varphi \right) $. The scalar prefactor in
our weak-field expansion is%
\begin{equation}
Z\left( l;\varphi ,\varphi ^{\prime }\right) =-\frac{1}{2}\alpha \left(
l\right) ^{2}\left( \sin ^{2}\left( \varphi \right) +\sin ^{2}\left( \varphi
^{\prime }\right) \right) \mathbf{1.}  \label{Zldefin1}
\end{equation}
Expressing $\sin ^{2}\left( \varphi \right) $ in terms of exponents, we can
write $Z\left( l;\varphi ,\varphi ^{\prime }\right) $ as a harmonic sum in $%
e^{is\varphi }$ (and similarly for $\varphi ^{\prime }$). The coefficient
multiplying the harmonic $\delta ^{\left( 4\right) }\left( p^{\prime
}-p-k-n\kappa \right) $ is%
\begin{equation}
Z_{s}\left( l\right) =\alpha \left( l\right) ^{2}z_{s}\mathbf{1,\;}z_{0}=-%
\frac{1}{2},\;z_{+2}=z_{-2}=+\frac{1}{4},  \label{Zslvol}
\end{equation}%
$z_{s}=0$ for all other $s$. Equivalently, written out:%
\begin{equation}
Z_{0}\left( l\right) =-\frac{1}{2}\alpha \left( l\right) ^{2}\mathbf{1,\;}%
Z_{\pm 2}\left( l\right) =+\frac{1}{4}\alpha _{l}^{2}\mathbf{1,}
\label{Z02lvol}
\end{equation}%
$Z_{s}\left( l\right) =0$ ($s\neq 0,\pm 2$). Here, $\mathbf{1}$ represents
the identity in Dirac and color space; $Zs$ is a scalar dressing factor. So%
\begin{eqnarray}
\Gamma _{\left( 2\right) ,s}^{c\mu }|_{Z} &=&\left( -ig^{2}\right) \int
\frac{d^{4}r}{\left( 2\pi \right) ^{4}}\frac{iH_{\alpha \beta }\left(
q\right) }{D_{P}D_{r}D_{g}}  \notag \\
&&\times \left[ \gamma ^{\alpha }T^{a}\left( i\left( \slashed{P}+m\right)
\right) Z_{s}\left( P\right) \gamma ^{\mu }T^{c}\left( i\left( \slashed{r}%
+m\right) \right) \gamma ^{\beta }T^{a}\right.  \notag \\
&&\left. +\gamma ^{\alpha }T^{a}\left( i\left( \slashed{P}+m\right) \right)
\gamma ^{\mu }T^{c}\left( i\left( \slashed{r}+m\right) \right) Z_{s}\left(
r\right) \gamma ^{\beta }T^{a}\right] ,  \label{Gamma2sZ}
\end{eqnarray}%
with $s=0$, $\pm 2$ only. So $\Gamma _{\left( 2\right) ,s}^{c\mu }\left(
p,k\right) =\Gamma _{\left( 2\right) ,s}^{c\mu }|_{\mathcal{O}^{\left(
1\right) }\mathcal{O}^{\left( 1\right) }}+\Gamma _{\left( 2\right) ,s}^{c\mu
}|_{\mathcal{O}^{\left( 2\right) }}+\Gamma _{\left( 2\right) ,s}^{c\mu
}|_{Z} $.

The full vertex in the periodic background decomposes as%
\begin{equation}
\Gamma ^{c\mu }\left( p^{\prime },p\right) =\sum_{n\in \mathcal{Z}}\left(
2\pi \right) ^{4}\delta ^{\left( 4\right) }\left( p^{\prime }-p-k-n\kappa
\right) \Gamma _{\left[ n\right] }^{c\mu }\left( p,k\right) .
\label{fullvertexperiod}
\end{equation}
where%
\begin{equation}
\Gamma _{\left[ \pm 1\right] }^{c\mu }=\Gamma _{\left( 1\right) ,\pm
1}^{c\mu },\;\Gamma _{\left[ 0\right] }^{c\mu }=\Gamma _{\left[ 0\right]
}^{c\mu }=\Gamma _{\left( 0\right) }^{c\mu }+\Gamma _{\left( 2\right)
,0}^{c\mu },\;\Gamma _{\left[ \pm 2\right] }^{c\mu }=\Gamma _{\left(
2\right) ,\pm 2}^{c\mu }.  \label{Gam102}
\end{equation}

Now, we examine the UV behavior of the one-loop correction. For large
Euclidean $r$ (or large Minkowski invariant $r^{2}$), the vacuum one-loop
vertex behaves as usual and is logarithmically UV divergent. Each linear
insertion carries a factor $U^{\left( 1\right) }\left( l\right) \sim \frac{g%
}{l\cdot \kappa }\slashed{\kappa}\slashed{\varepsilon}$ $\Rightarrow $ $%
U^{\left( 1\right) }\left( l\right) \sim \mathcal{O}\left( \frac{1}{r}%
\right) $ ($r\rightarrow \infty $), because $l\cdot \kappa \sim r$ for
generic directions. Quadratic pieces scale as $U^{\left( 2\right) }\left(
l\right) \sim \mathcal{O}\left( \frac{1}{\left( l\cdot \kappa \right) ^{2}}%
\right) =\mathcal{O}\left( \frac{1}{r^{2}}\right) $. As a result, $\Gamma
_{(0)}$ has the same UV divergence as the vacuum QCD vertex. The integral is
UV-convergent since we obtain an extra factor of $1/r$ when comparing $%
\Gamma _{\left( 1\right) }$ to the vacuum. The convergence of $\Gamma
_{\left( 2\right) }$ is increased since we obtain $1/r^{2}$ (or two factors
of $1/r$). So, the UV divergent part is entirely contained within the
vacuum-like component: $\left[ \Gamma ^{c\mu }\right] _{\mathrm{div}}=\left[
\Gamma _{\left( 0\right) }^{c\mu }\right] _{\mathrm{div}}$,$\;\left[ \Gamma
_{\left( 1\right) }^{c\mu }\right] _{\mathrm{div}}=0$,$\;\left[ \Gamma
_{\left( 2\right) }^{c\mu }\right] _{\mathrm{div}}=0$, up to possible
scheme-dependent subtleties with the axial-gauge prescription (which
influence how you handle spurious poles but do not introduce new UV
divergences related to the background). Since $\Gamma _{\left( 1\right) }$
and $\Gamma _{\left( 2\right) }$ are UV finite, the background only affects
the finite remainder: $\Gamma ^{c\mu }=\Gamma _{\left( 0\right) }^{c\mu
}+\Gamma _{\left( 1\right) }^{c\mu }|_{finite}+\Gamma _{\left( 2\right)
}^{c\mu }|_{finite}+\cdots $, and those finite pieces come with harmonic
delta functions that enforce $p^{\prime }-p-k=n\kappa $.

\section{Fermion effective mass}

The fermion self-energy is given by%
\begin{equation}
\Sigma \left( p\right) =i\int \frac{d^{4}q}{\left( 2\pi \right) ^{4}}\left(
-ig\gamma ^{\mu }T^{a}\right) \tilde{G}\left( p-q\right) \left( -ig\gamma
^{\nu }T^{b}\right) \tilde{D}_{\mu \nu }^{ab}\left( q\right)
\label{fermionselfernergydef}
\end{equation}

Substituting $\tilde{G}$ and $\tilde{D}_{\mu \nu }^{ab}$ into (\ref%
{fermionselfernergydef}), we obtain%
\begin{equation}
\Sigma \left( p\right) =-ig^{2}\int \frac{d^{4}q}{\left( 2\pi \right) ^{4}}%
\left\{ \mathcal{I}_{1}-\mathcal{I}_{2}-\mathcal{I}_{3}\right\} ,
\label{fermionWdef}
\end{equation}%
\ where%
\begin{eqnarray}
\mathcal{I}_{1} &=&\frac{\gamma ^{\mu }T^{a}\left( \slashed{r}+m\right)
U\left( r\right) \gamma _{\mu }T^{a}}{D_{F}\left( r\right) D_{G}\left(
q\right) },  \notag \\
\mathcal{I}_{2} &=&\frac{\slashed{q}T^{a}\left( \slashed{r}+m\right) U\left(
r\right) \slashed{n}T^{a}\left( n^{\ast }\cdot q\right) }{D_{F}\left(
r\right) D_{G}\left( q\right) D_{ax}\left( q\right) },  \notag \\
\mathcal{I}_{3} &=&\frac{\slashed{n}T^{a}\left( \slashed{r}+m\right) U\left(
r\right) \slashed{q}T^{a}\left( n^{\ast }\cdot q\right) }{D_{F}\left(
r\right) D_{G}\left( q\right) D_{ax}\left( q\right) },  \notag \\
D_{F}\left( r\right) &\equiv &\left( r^{2}-m^{2}+i\epsilon \right)
,\;D_{G}\left( q\right) \equiv \left( q^{2}+i\epsilon \right) ,  \notag \\
D_{ax}\left( q\right) &\equiv &\left( n^{\ast }\cdot q\right) \left( q\cdot
n\right) +i\epsilon ,\;r\equiv p-q,  \label{Wdef}
\end{eqnarray}%
We use the cyclicity of the trace: $\mathrm{Tr}\left[ \left( \slashed{p}%
+m\right) \gamma ^{\mu }L\gamma _{\mu }\right] =\mathrm{Tr}\left[ \left( -2%
\slashed{p}+4m\right) L\right] $, and define%
\begin{equation}
\delta m\equiv \frac{1}{4m}\mathrm{Tr}\left[ \left( \slashed{p}+m\right)
\Sigma \left( p\right) \right] |_{p^{2}=m^{2}},  \label{defdeltam}
\end{equation}%
so that (to one-loop order) $m_{eff}=m+\delta m$. It is clear that as the
limit $\mathcal{A\rightarrow }0$, $U\rightarrow 1$, and we substitute $%
T^{a}=1$, $g=e$, we obtain%
\begin{equation}
\Sigma _{El}\left( p\right) =-ie^{2}\int \frac{d^{4}q}{\left( 2\pi \right)
^{4}}\left[ \frac{\gamma ^{\mu }\left( \slashed{r}+m\right) \gamma _{\mu }}{%
D_{F}\left( r\right) D_{G}\left( q\right) }-\frac{\slashed{q}\left( %
\slashed{r}+m\right) \slashed{n}\left( n^{\ast }\cdot q\right) }{D_{F}\left(
r\right) D_{G}\left( q\right) D_{ax}\left( q\right) }-\frac{\slashed{n}%
\left( \slashed{r}+m\right) \slashed{q}\left( n^{\ast }\cdot q\right) }{%
D_{F}\left( r\right) D_{G}\left( q\right) D_{ax}\left( q\right) }\right] .
\label{elecselfenergy}
\end{equation}%
This corresponds to the electron's self-energy in electromagnetic
interactions.

We expand $U(r)$, keeping $C\left( r\right) \equiv \cos \left( \theta \left(
p,\varphi \right) \right) \cos \left( \theta \left( p,\varphi ^{\prime
}\right) \right) $,
\begin{equation}
U\left( r\right) =C\left( r\right) \left[ 1+\Delta ^{\left( 1\right) }\left(
r\right) +\Delta ^{\left( 2\right) }\left( r\right) \right] ,
\label{Uexpand}
\end{equation}
where%
\begin{eqnarray}
\alpha ^{\prime } &\equiv &\frac{g}{2\left( r\cdot \kappa \right) }\frac{%
\tan \left( \theta \left( p,\varphi ^{\prime }\right) \right) }{\theta
\left( p,\varphi ^{\prime }\right) },\;\beta \equiv \frac{g}{2\left( r\cdot
\kappa \right) }\frac{\tan \left( \theta \left( p,\varphi \right) \right) }{%
\theta \left( p,\varphi \right) },  \notag \\
\Delta ^{\left( 1\right) }\left( r\right) &=&\alpha ^{\prime }\mathcal{%
\slashed{A}}^{e}\left( \varphi ^{\prime }\right) \slashed{\kappa}T_{e}+\beta %
\slashed{\kappa}\mathcal{\slashed{A}}^{b}\left( \varphi \right) T_{b},
\notag \\
\Delta ^{\left( 2\right) }\left( r\right) &=&\alpha ^{\prime }\beta \left(
\mathcal{\slashed{A}}^{e}\left( \varphi ^{\prime }\right) \slashed{\kappa}%
\right) \left( \slashed{\kappa}\mathcal{\slashed{A}}^{b}\left( \varphi
\right) \right) T_{b}T_{e}.  \label{Delta1Delta2def}
\end{eqnarray}%
We have for the "1" in $U$, $T^{a}\left( \cdots \right) T^{a}\rightarrow
C_{F}\left( \cdots \right) $ with $C_{F}=\frac{N^{2}-1}{2N}$ (in the
fundamental representation of $SU(N)$). For $\delta m_{\left( 1\right) }$ we
have $\delta m_{\left( 1\right) }\equiv \frac{1}{4m}\mathrm{Tr}\left[ \left( %
\slashed{p}+m\right) \gamma ^{\mu }\left( \slashed{r}+m\right) U\left(
r\right) \gamma _{\mu }\right] =\frac{1}{4m}\mathrm{Tr}\left[ \left( -2%
\slashed{p}+4m\right) \left( \slashed{r}+m\right) U\left( r\right) \right] $%
. Replacing $U(r)$, we get for a vacuum-like segment (no explicit $\mathcal{A%
}$ insertion beyond $C(r)$) the expression:%
\begin{equation}
\delta m_{\left( 1\right) }^{\left( 0\right) }=C_{F}C\left( r\right) \left(
4m-\frac{2}{m}p\cdot r\right)  \label{deltamvac1}
\end{equation}%
then we compute linear-in-$\mathcal{A}$ insertions from $\Delta ^{\left(
1\right) }$. For $\mathcal{\slashed{A}}^{e}\left( \varphi ^{\prime }\right) %
\slashed{\kappa}$, and $\slashed{\kappa}\mathcal{\slashed{A}}^{b}$ using $%
k^{2}=0$, $\kappa \mathcal{A}^{e}\left( \varphi ^{\prime }\right) =0$, $%
\kappa \mathcal{A}^{b}\left( \varphi \right) =0$.%
\begin{eqnarray}
\frac{1}{4m}\mathrm{Tr}\left[ \left( -2\slashed{p}+4m\right) \left( %
\slashed{r}+m\right) \mathcal{\slashed{A}}^{e}\left( \varphi ^{\prime
}\right) \slashed{\kappa}\right] &=&\frac{2}{m}\left[ \left( p\cdot \mathcal{%
A}^{e}\left( \varphi ^{\prime }\right) \right) \left( r\cdot \kappa \right)
-\left( p\cdot \kappa \right) \left( r\cdot \mathcal{A}^{e}\left( \varphi
^{\prime }\right) \right) \right] ,  \notag \\
\frac{1}{4m}\mathrm{Tr}\left[ \left( -2\slashed{p}+4m\right) \left( %
\slashed{r}+m\right) \slashed{\kappa}\mathcal{\slashed{A}}^{b}\left( \varphi
\right) \right] &=&\frac{2}{m}\left[ \left( p\cdot \kappa \right) \left(
r\cdot \mathcal{A}^{b}\left( \varphi \right) \right) -\left( p\cdot \mathcal{%
A}^{b}\left( \varphi \right) \right) \left( r\cdot \kappa \right) \right] .
\label{Tr(kA)}
\end{eqnarray}%
So the linear part of $\delta m_{\left( 1\right) }$ is%
\begin{eqnarray}
\delta m_{\left( 1\right) }^{\left( 1\right) } &=&C\left( r\right) \left(
C_{F}-\frac{C_{A}}{2}\right) \left[ \alpha ^{\prime }\frac{2}{m}\left(
\left( p\cdot \mathcal{A}^{e}\left( \varphi ^{\prime }\right) \right) \left(
r\cdot \kappa \right) -\left( p\cdot \kappa \right) \left( r\cdot \mathcal{A}%
^{e}\left( \varphi ^{\prime }\right) \right) \right) T_{e}\right.  \notag \\
&&\left. +\beta \frac{2}{m}\left( \left( p\cdot \kappa \right) \left( r\cdot
\mathcal{A}^{b}\left( \varphi \right) \right) -\left( p\cdot \mathcal{A}%
^{b}\left( \varphi \right) \right) \left( r\cdot \kappa \right) \right) T_{b}%
\right] .  \label{deltam11}
\end{eqnarray}%
where we used $T^{a}T^{b}T^{a}=\left( C_{F}-\frac{C_{A}}{2}\right) T^{b}$,$%
\;C_{A}=N$. For the vacuum-like axial trace (again exact in $C\left(
r\right) $) with $U\rightarrow C\left( r\right) $ and color $T^{a}\left(
\cdots \right) T^{a}\rightarrow C_{F}\left( \cdots \right) $, we have the
linear axial trace from $\Delta ^{\left( 1\right) }$ (for the $\alpha
^{\prime }\mathcal{\slashed{A}}^{e}\left( \varphi ^{\prime }\right) %
\slashed{k}T_{e}$ insertion):%
\begin{eqnarray}
\delta m_{\left( 2+3\right) }^{\left( 1\right) }|_{\mathcal{A}^{\prime }}
&=&C\left( r\right) \alpha ^{\prime }\frac{2}{m}\left[ -\left( \mathcal{A}%
^{e}\left( \varphi ^{\prime }\right) \cdot n\right) \left( p\cdot q\right)
\right.  \notag \\
&&\left. +\left( p\cdot \mathcal{A}^{e}\left( \varphi ^{\prime }\right)
\right) \left( r\cdot \kappa \right) \left( n\cdot q\right) -\left( \mathcal{%
A}^{e}\left( \varphi ^{\prime }\right) \cdot q\right) \left( r\cdot \kappa
\right) \left( n\cdot p\right) \right.  \notag \\
&&\left. +\left( \mathcal{A}^{e}\left( \varphi ^{\prime }\right) \cdot
r\right) \left( \kappa \cdot n\right) \left( p\cdot q\right) -\left(
\mathcal{A}^{e}\left( \varphi ^{\prime }\right) \cdot r\right) \left( \kappa
\cdot p\right) \left( n\cdot q\right) \right.  \notag \\
&&\left. +\left( \mathcal{A}^{e}\left( \varphi ^{\prime }\right) \cdot
r\right) \left( \kappa \cdot q\right) \left( n\cdot p\right) \right] \left(
C_{F}-\frac{C_{A}}{2}\right) T_{e},  \label{deltam123alfsh}
\end{eqnarray}%
(for the $\beta \mathcal{\slashed{A}}^{e}\left( \varphi ^{\prime }\right) %
\slashed{k}T_{b}$ insertion):%
\begin{eqnarray}
\delta m_{\left( 2+3\right) }^{\left( 1\right) }|_{\mathcal{A}} &=&C\left(
r\right) \beta \frac{2}{m}\left[ +\left( \mathcal{A}^{b}\left( \varphi
\right) \cdot n\right) \left( r\cdot \kappa \right) \left( p\cdot q\right)
\right.  \notag \\
&&\left. -\left( p\cdot \mathcal{A}^{b}\left( \varphi \right) \right) \left(
r\cdot \kappa \right) \left( n\cdot q\right) +\left( \mathcal{A}^{b}\left(
\varphi \right) \cdot q\right) \left( r\cdot \kappa \right) \left( n\cdot
p\right) \right.  \notag \\
&&\left. -\left( \mathcal{A}^{b}\left( \varphi \right) \cdot r\right) \left(
\kappa \cdot n\right) \left( p\cdot q\right) +\left( \mathcal{A}^{b}\left(
\varphi \right) \cdot r\right) \left( \kappa \cdot p\right) \left( n\cdot
q\right) \right.  \notag \\
&&\left. -\left( \mathcal{A}^{b}\left( \varphi \right) \cdot q\right) \left(
r\cdot \kappa \right) \left( n\cdot p\right) \right] \left( C_{F}-\frac{C_{A}%
}{2}\right) T_{b}.  \label{deltam123bet}
\end{eqnarray}%
Under the same plane-wave constraints $k^{2}=0$, $k\mathcal{A}^{b}\left(
\varphi \right) =0$, $k\mathcal{A}^{e}\left( \varphi ^{\prime }\right) =0$,
the scalar mass projector from $\Delta ^{\left( 2\right) }$ also vanishes: $%
\delta m_{\left( 2+3\right) }^{\left( 2\right) }=0$. Collecting the traced
results, the complete one-loop on-shell mass shift is%
\begin{eqnarray}
\delta m &=&-ig^{2}\int \frac{d^{4}q}{\left( 2\pi \right) ^{4}}\frac{1}{%
D_{F}\left( r\right) D_{G}\left( q\right) }  \notag \\
&&\times \left\{ \delta m_{\left( 1\right) }^{\left( 0\right) }+\delta
m_{\left( 1\right) }^{\left( 1\right) }-\left[ -\frac{\left( n^{\ast }\cdot
q\right) }{D_{ax}\left( q\right) }\right] \left( \delta m_{\left( 2+3\right)
}^{\left( 0\right) }+\delta m_{\left( 2+3\right) }^{\left( 1\right) }\right)
\right\} .  \label{deltamfull}
\end{eqnarray}

For transverse plane wave the phase integral provides $\int_{0}^{\varphi
}d\varphi ^{\prime }\left( \mathcal{A}^{a}\cdot r\right) =\left( \varepsilon
^{a}\cdot r\right) \sin \left( \varphi \right) $ and for $\theta \left(
r,\varphi \right) $, we have $\theta \left( r,\varphi \right) =\alpha \left(
r\right) \sin \left( \varphi \right) $, where $\alpha \left( r\right) =\frac{%
g}{r\cdot \kappa }\sqrt{\frac{\left( \varepsilon ^{a}\cdot r\right) \left(
\varepsilon _{a}\cdot r\right) }{2N}}$ and
\begin{equation}
\langle \cos \left( \alpha \left( r\right) \varphi \right) \rangle _{\varphi
}=\frac{1}{2\pi }\int_{0}^{2\pi }d\varphi \cos \left( \alpha \left( r\right)
\sin \left( \varphi \right) \right) =J_{0}\left( \alpha \left( r\right)
\right) ,  \label{cosaver}
\end{equation}%
\begin{equation}
\langle C\left( r\right) \rangle _{\varphi ,\varphi ^{\prime }}=\langle \cos
\left( \alpha \left( r\right) \varphi \right) \rangle _{\varphi
}^{2}=J_{0}^{2}\left( \alpha \left( r\right) \right) .  \label{<C>}
\end{equation}%
Using the tracing results obtained above, the cycle-averaged mass
displacement on the shell surface is%
\begin{equation}
\langle \delta m\rangle =-ig^{2}C_{F}\int \frac{d^{4}q}{\left( 2\pi \right)
^{4}}\frac{\mathcal{W}\left( p-q\right) \left[ \mathcal{L}_{F}\left(
p,q\right) +\mathcal{L}_{ML}\left( p,q;n,n^{\ast }\right) \right] }{\left[
\left( p-q\right) ^{2}-m^{2}+i\epsilon \right] \left( q^{2}+i\epsilon
\right) }  \label{deltamass}
\end{equation}%
where $\mathcal{W}\left( r\right) =J_{0}^{2}\left( \alpha \left( r\right)
\right) $. The traced numerators (vacuum-like pieces) are $\mathcal{L}%
_{F}\left( p,q\right) =2m+\frac{2}{m}\left( p\cdot q\right) $, and the
ML/axial term (from our $\mathcal{I}_{2}+\mathcal{I}_{3}$) is%
\begin{eqnarray}
\mathcal{L}_{ML}\left( p,q;n,n^{\ast }\right) &=&\frac{n^{\ast }\cdot q}{%
\left( n^{\ast }\cdot q\right) \left( n\cdot q\right) +i\epsilon }\frac{2}{m}
\notag \\
&&\times \left[ m^{2}\left( n\cdot q\right) -\left( m^{2}-p\cdot q\right)
\left( q\cdot n\right) \right.  \notag \\
&&\left. +\left( p\cdot n\right) \left( p\cdot q\right) +\left( p\cdot
q\right) \left( p\cdot n-q\cdot n\right) \right] .  \label{NML}
\end{eqnarray}

To apply the Wick rotation, we can express the loop energy as $%
q^{0}\rightarrow iq_{4\text{,}}\;d^{4}q\rightarrow id^{4}q_{E}$,$%
\;q^{2}=\left( q^{0}\right) ^{2}-\mathbf{q}^{2}\rightarrow -q_{E}^{2}$. So,
the Euclidean form is%
\begin{equation}
\langle \delta m\rangle =g^{2}C_{F}\int \frac{d^{4}q_{E}}{\left( 2\pi
\right) ^{4}}\frac{\mathcal{W}\left( r_{E}\right) \left[ \mathcal{L}%
_{F}\left( p_{E},q_{E}\right) +\mathcal{L}_{ML}\left(
p_{E},q_{E};n_{E},n_{E}^{\ast }\right) \right] }{q_{E}^{2}\left( \left(
p_{E}-q_{E}\right) ^{2}+m^{2}\right) },  \label{Euclform}
\end{equation}%
where%
\begin{eqnarray}
\mathcal{W}\left( r_{E}\right) &=&J_{0}^{2}\left( \alpha _{E}\left(
r_{E}\right) \right) ,  \notag \\
\alpha _{E}\left( r_{E}\right) &=&\frac{g}{r_{E}\kappa _{E}}\sqrt{\frac{%
\left( \varepsilon ^{a}\cdot r_{E}\right) \left( \varepsilon _{a}\cdot
r_{E}\right) }{2N}},\;\kappa _{E}\varepsilon ^{a}=0,\;\kappa _{E}^{2}=0,
\label{WEucl}
\end{eqnarray}%
\begin{equation}
\mathcal{L}_{F,E}\left( p_{E},q_{E}\right) =4m-\frac{2}{m}p_{E}\cdot r_{E},
\label{LFEucl}
\end{equation}%
\begin{eqnarray}
\mathcal{L}_{ML,E}\left( p_{E},q_{E};n_{E},n_{E}^{\ast }\right) &=&\frac{%
n_{E}^{\ast }\cdot q_{E}}{\left( n_{E}^{\ast }\cdot q_{E}\right) \left(
n_{E}\cdot q_{E}\right) }\frac{2}{m}  \notag \\
&&\times \left[ m^{2}\left( n_{E}\cdot q_{E}\right) -\left( m^{2}-p_{E}\cdot
q_{E}\right) \left( q_{E}\cdot n_{E}\right) \right.  \notag \\
&&\left. +\left( p_{E}\cdot n_{E}\right) \left( p_{E}\cdot q_{E}\right)
+\left( p_{E}\cdot q_{E}\right) \left( p_{E}\cdot n_{E}-q_{E}\cdot
n_{E}\right) \right] .  \label{LMLEucl}
\end{eqnarray}%
We introduce the Feynman parameter $x\in \left[ 0,1\right] $ to combine the
denominators:%
\begin{eqnarray}
&&\frac{1}{q_{E}^{2}\left( \left( p_{E}-q_{E}\right) ^{2}+m^{2}\right) }
\notag \\
&=&\int_{0}^{1}dx\frac{1}{\left[ \left( q_{E}-xp_{E}\right) ^{2}+x\left(
1-x\right) p_{E}^{2}+xm^{2}\right] ^{2}}.  \label{xdenominat}
\end{eqnarray}%
Denoting $l_{E}=q_{E}-xp_{E}$, for $r_{E}$ we have $r_{E}=p_{E}-q_{E}=\left(
1-x\right) p_{E}-l_{E}$. Then%
\begin{equation}
\langle \delta m\rangle =g^{2}C_{F}\int_{0}^{1}dx\int \frac{d^{4}l_{E}}{%
\left( 2\pi \right) ^{4}}\frac{\mathcal{W}\left( \left( 1-x\right)
p_{E}-l_{E}\right) \left[ \mathcal{\tilde{L}}_{F,E}+\mathcal{\tilde{L}}%
_{ML,E}\right] }{\left( l_{E}^{2}+\Delta \left( x\right) \right) ^{2}}
\label{<deltamEuc>}
\end{equation}%
with $\Delta \left( x\right) =x\left( 1-x\right) p_{E}^{2}+xm^{2}$. and on
shell $p_{E}^{2}\rightarrow -m^{2}$ so $\Delta \left( x\right) =x^{2}m^{2}$.
At a large Euclidean loop momentum $\alpha _{E}\left( r_{E}\right) \sim
\frac{1}{r_{E}\cdot \kappa _{E}}\times \left( \varepsilon \cdot r_{E}\right)
\sim \mathcal{O}\left( 1\right) $, the Bessel factor remains bounded, and in
practice, we can safely split: $\langle \delta m\rangle _{E}=\delta
m_{E}^{free}+\delta m_{E}^{bg\left( finit\right) }$, where $\delta
m_{E}^{free}$ is the standard one-loop mass shift, and $\delta
m_{E}^{bg\left( finit\right) }$ is derived from the difference integral%
\begin{equation}
\delta m_{E}^{bg\left( finit\right) }=g^{2}C_{F}\int_{0}^{1}dx\int \frac{%
d^{4}l_{E}}{\left( 2\pi \right) ^{4}}\frac{\left[ \mathcal{W}\left( r\left(
l_{E},x\right) \right) -1\right] }{\left( l_{E}^{2}+x^{2}m^{2}\right) ^{2}}%
\left[ \mathcal{\tilde{L}}_{F,E}\left( l_{E},x\right) +\mathcal{\tilde{L}}%
_{ML,E}\left( l_{E},x\right) \right] .  \label{deltamEbg}
\end{equation}%
The background-dependent correction is finite because $\mathcal{W}\left(
r\left( l_{E},x\right) \right) -1$ kills the local UV piece. The small-field
expansion is $\mathcal{W}\left( r\left( l_{E},x\right) \right) -1=-\frac{1}{2%
}\alpha _{E}\left( r_{E}\right) ^{2}+\mathcal{O}\left( \alpha
_{E}^{4}\right) ,$ so to $O\left( \varepsilon ^{4}\right) $,%
\begin{equation}
\delta m_{E}^{\left( 2\right) }=-\frac{1}{2}g^{2}C_{F}\int_{0}^{1}dx\int
\frac{d^{4}l_{E}}{\left( 2\pi \right) ^{4}}\frac{\alpha _{E}\left(
r_{E}\right) ^{2}}{\left( l_{E}^{2}+x^{2}m^{2}\right) ^{2}}\left[ \mathcal{%
\tilde{L}}_{F,E}\left( l_{E},x\right) +\mathcal{\tilde{L}}_{ML,E}\left(
l_{E},x\right) \right] .  \label{deltamE2}
\end{equation}%
All background dependence appears through UV-finite factors, such as $\cos
\left( \theta \right) $, $J_{0}\left( \alpha \right) ^{2}$, and $\left(
\varepsilon \cdot p\right) ^{2}/\left( p\cdot \kappa \right) ^{2}$,
confirming that mass renormalization remains local. Therefore, external
classical fields do not alter the UV structure of renormalizable QFT.

\section{Fermion condensate in an external Yang-Mills gauge field}

\label{Fermioncond}

The field-induced condensate is computed by explicitly subtracting $\langle
0|\bar{\psi}\psi |0\rangle _{\mathcal{A}}-\langle 0|\bar{\psi}\psi |0\rangle
_{free}$, which removes the divergence of the free propagator. This method
aligns with renormalization schemes where free-field subtractions regularize
VEVs before incorporating interaction effects. We write

\begin{equation}
\langle 0|\bar{\psi}\psi |0\rangle _{\mathcal{A-}free}=\langle 0|\bar{\psi}%
\psi |0\rangle _{\mathcal{A}}-\langle 0|\bar{\psi}\psi |0\rangle _{free}
\label{normalordered}
\end{equation}%
where\
\begin{equation}
\langle \bar{\psi}\psi \rangle _{\mathcal{A}}=-\int \frac{d^{4}p}{\left(
2\pi \right) ^{4}}\mathrm{Tr}\left[ \tilde{G}_{F}\left( p\right) \right] ,
\label{ferconddef}
\end{equation}

Substituting (\ref{Fermionpropmomspace}) into (fer\ref{ferconddef}) with $%
x=x^{\prime }$, we obtain%
\begin{equation}
\langle \bar{\psi}\psi \rangle _{\mathcal{A}}=-\int \frac{d^{4}p}{\left(
2\pi \right) ^{4}}\mathrm{Tr}\left[ \frac{i\left( \slashed{p}+m\right)
\tilde{U}\left( p\right) }{p^{2}-m^{2}+i\epsilon }\right] ,  \label{fercondU}
\end{equation}%
where
\begin{eqnarray}
&&U\left( p\right) |_{x=x^{\prime }}  \notag \\
&=&\tilde{U}\left( p\right) =\cos ^{2}\left( \theta \left( p,\varphi \right)
\right)  \notag \\
&&\times \left\{ 1+g\frac{\tan \left( \theta \left( p,\varphi \right)
\right) }{2\left( p\kappa \right) \theta \left( p,\varphi \right) }\left(
\left( \left( \gamma ^{\sigma }\right) ^{\dagger }\mathcal{A}_{\sigma
}^{e}\left( \varphi \right) \right) \left( \left( \gamma ^{\rho }\right)
^{\dagger }k_{\rho }\right) T_{e}+\left( \gamma ^{\nu }k_{\nu }\right)
\left( \gamma ^{\lambda }\mathcal{A}_{\lambda }^{b}\left( \varphi \right)
\right) T_{b}\right) \right.  \notag \\
&&\left. +g^{2}\left( \frac{\tan \left( \theta \left( p,\varphi \right)
\right) }{2\left( p\kappa \right) \theta \left( p,\varphi \right) }\right)
^{2}\left( \gamma ^{\nu }k_{\nu }\right) \left( \gamma ^{\lambda }\mathcal{A}%
_{\lambda }^{b}\left( \varphi \right) \right) \left( \left( \gamma ^{\sigma
}\right) ^{\dagger }\mathcal{A}_{\sigma }^{e}\left( \varphi \right) \right)
\left( \left( \gamma ^{\rho }\right) ^{\dagger }k_{\rho }\right)
T_{b}T_{e}\right\} ,  \label{Ux}
\end{eqnarray}%
and
\begin{equation}
\theta \left( p\right) =\theta \left( p,\varphi \right) =\frac{g}{\left(
p\kappa \right) }\sqrt{\frac{1}{2N}}\left( \int_{0}^{\varphi }d\varphi
^{\prime }\left( \mathcal{A}_{\mu }^{a}\left( \varphi ^{\prime }\right)
p^{\mu }\right) \int_{0}^{\varphi }d\varphi ^{\prime \prime }\left( \mathcal{%
A}_{a}^{\mu }\left( \varphi ^{\prime \prime }\right) p_{\mu }\right) \right)
^{\frac{1}{2}},  \label{tetx}
\end{equation}%
When $U(p)\rightarrow 1$, we obtain%
\begin{equation}
\mathrm{Tr}\left[ i\left( \slashed{p}+m\right) \right] =i\mathrm{Tr}%
_{D}\left( \slashed{p}+m\right) \mathrm{Tr}_{c}\left( \mathbf{1}\right)
=i\left( 0+4m\right) N=4imN  \label{trfree}
\end{equation}%
so%
\begin{equation}
\langle \bar{\psi}\psi \rangle _{free}=-4imN\int \frac{d^{4}p}{\left( 2\pi
\right) ^{4}}\frac{1}{p^{2}-m^{2}+i\epsilon }  \label{condfree}
\end{equation}%
This is UV divergent and must be regularized.

Considering that only terms with an even number of gamma matrices contribute
and $\mathrm{tr}_{c}\left( T^{a}\right) =0$, and in axial gauge with $\kappa
^{\mu }\mathcal{A}_{\mu }^{a}=0$ and $\kappa ^{2}=0$, we have%
\begin{eqnarray}
\langle 0|\bar{\psi}\psi |0\rangle _{\mathcal{A-}free} &=&-4imN\int \frac{%
d^{4}p}{\left( 2\pi \right) ^{4}}\frac{\cos ^{2}\left( \theta \left(
p,\varphi \right) \right) -1}{p^{2}-m^{2}+i\epsilon }  \notag \\
&=&2imN\int \frac{d^{4}p}{\left( 2\pi \right) ^{4}}\frac{1-\cos \left(
2\theta \left( p,\varphi \right) \right) }{p^{2}-m^{2}+i\epsilon },
\label{fercondcos}
\end{eqnarray}

The condensate vanishes identically when the background is switched off ($%
\theta \rightarrow 0$). So this is a genuine response functional of the
fermionic vacuum to the background gauge field. The dependence on the
background enters only through the gauge-invariant scalar phase $\theta
\left( p,\varphi \right) $. The contribution of field-free divergence is
removed through subtraction. The integrand has a clear sign in Euclidean
space because $1-\cos (2\theta )=2\sin ^{2}\theta \geq 0$. The background
always makes the scalar density bigger than it would be without the field.

In the case of the monochromatic plane wave, it is useful to package the
quadratic form%
\begin{equation}
\left( \varepsilon ^{a}\cdot p\right) \left( \varepsilon _{a}\cdot p\right)
=p_{\mu }\Xi ^{\mu \nu }p_{\nu },\;\Xi ^{\mu \nu }\equiv \varepsilon ^{a\mu
}\varepsilon _{a}^{\nu }.  \label{epep}
\end{equation}%
So%
\begin{equation}
2\theta \left( p,\varphi \right) =\frac{2g}{p\kappa }\sqrt{\frac{1}{2N}}%
\sqrt{p_{\mu }\Xi ^{\mu \nu }p_{\nu }}\sin \left( \varphi \right) .
\label{2tetpfi}
\end{equation}%
For a monochromatic wave, the clearest \textquotedblleft oscillatory
contribution\textquotedblright\ is the average over $\varphi \in \left[
0,2\pi \right] $. We use the exact Bessel identity $\frac{1}{2\pi }%
\int_{0}^{2\pi }d\varphi \cos \left( a\sin \varphi \right) =J_{0}\left(
a\right) $. Then%
\begin{equation}
\langle \cos \left( 2\theta \left( p,\varphi \right) \right) \rangle
_{\varphi }=J_{0}\left( \frac{2g}{p\kappa }\sqrt{\frac{1}{2N}}\sqrt{p\Xi p}%
\right) .  \label{cosaverage}
\end{equation}%
Therefore, the precise phase-averaged condensate is%
\begin{equation}
\langle \langle 0|\bar{\psi}\psi |0\rangle _{\mathcal{A-}free}\rangle
_{\varphi }=2imN\int \frac{d^{4}p}{\left( 2\pi \right) ^{4}}\frac{%
1-J_{0}\left( \frac{2g}{p\kappa }\sqrt{\frac{1}{2N}}\sqrt{p\Xi p}\right) }{%
p^{2}-m^{2}+i\epsilon }.  \label{exactphasavercond}
\end{equation}

In $d=4-2\eta $, we define the regulated (phase-averaged) integral:%
\begin{equation}
\langle \langle 0|\bar{\psi}\psi |0\rangle _{\mathcal{A-}free}\rangle
_{\varphi }^{DR}=2imN\mu ^{2\eta }\int \frac{d^{4-2\eta }p}{\left( 2\pi
\right) ^{4-2\eta }}\frac{1-J_{0}\left( \frac{2g}{p\kappa }\sqrt{\frac{1}{2N}%
}\sqrt{p\Xi p}\right) }{p^{2}-m^{2}+i\epsilon }.  \label{DRavercond}
\end{equation}

Pauli--Villars (PV) implements%
\begin{equation}
\frac{1}{p^{2}-m^{2}+i\epsilon }\rightarrow \frac{1}{p^{2}-m^{2}+i\epsilon }-%
\frac{1}{p^{2}-M^{2}+i\epsilon },\;M\gg m,  \label{PVreg}
\end{equation}%
so the regulated, phase-averaged condensate is%
\begin{eqnarray}
&&\langle \langle 0|\bar{\psi}\psi |0\rangle _{\mathcal{A-}free}\rangle
_{\varphi }^{PV}  \notag \\
&=&2imN\int \frac{d^{4}p}{\left( 2\pi \right) ^{4}}\left( \frac{1}{%
p^{2}-m^{2}+i\epsilon }-\frac{1}{p^{2}-M^{2}+i\epsilon }\right) \left[
1-J_{0}\left( \frac{2g}{p\kappa }\sqrt{\frac{1}{2N}}\sqrt{p\Xi p}\right) %
\right] .  \label{PVaverregul}
\end{eqnarray}

Now we choose the coordinates of the light front:%
\begin{equation}
p^{\pm }=p^{0}\pm p^{3},\;p^{2}=p^{+}p^{-}-p_{\perp }^{2}.
\label{lightfroncoor}
\end{equation}%
and%
\begin{equation}
\kappa ^{\mu }=\left( \varpi ,0,\mathbf{0}_{\perp }\right) \;\Rightarrow
\;p\kappa =\frac{1}{2}\varpi p^{-}.  \label{lightfrontmom}
\end{equation}%
In axial gauge for a plane wave, we choose $\varepsilon ^{a}$ purely
transverse, so%
\begin{equation}
\left( \varepsilon ^{a}\cdot p\right) =-\varepsilon _{\perp }^{a}\cdot
p_{\perp },\;\Rightarrow \;p\Xi p=p_{\perp }^{i}C_{ij}p_{\perp
}^{j},\;C_{ij}\equiv \varepsilon _{\perp }^{ai}\varepsilon _{\perp }^{aj}.
\label{Cdef}
\end{equation}%
We begin by integrating $p^{-}$ over residues using the phase-averaged
Minkowski form:%
\begin{equation}
\int^{+}\frac{dp^{-}}{2\pi }\frac{i}{p^{+}p^{-}-p_{\perp
}^{2}-m^{2}+i\epsilon }=\frac{1}{p^{+}}\theta \left( p^{+}\right) \frac{1}{2}%
,  \label{integrpmin}
\end{equation}%
where theta is the Heaviside function. The condensate then becomes%
\begin{equation}
\langle \langle 0|\bar{\psi}\psi |0\rangle _{\mathcal{A-}free}\rangle
_{\varphi }=mN\int_{p^{+}>0}\frac{dp^{+}d^{2}p_{\perp }}{\left( 2\pi \right)
^{3}}\frac{1}{p^{+}}\left[ 1-J_{0}\left( \frac{4g}{\varpi }\sqrt{\frac{1}{2N}%
}\frac{\sqrt{p_{\perp }Cp_{\perp }}}{p_{on}^{-}}\right) \right] ,
\label{condlightfront}
\end{equation}%
where%
\begin{equation}
p_{on}^{-}=\frac{p_{\perp }^{2}+m^{2}}{p^{+}}.  \label{pminuson}
\end{equation}%
We then use different light-front UV cutoff surfaces: \textquotedblleft
rectangular\textquotedblright\ cutoffs in $\left( p^{+},p_{\perp }\right) $
and covariant \textquotedblleft invariant mass\textquotedblright\
(Brodsky--Lepage) cutoff. We select the following conditions for
\textquotedblleft rectangular\textquotedblright\ cutoffs $\delta \leq
p^{+}\leq \Lambda ^{+},\ |p_{\perp }|\leq \Lambda _{\perp }$. Then
\begin{eqnarray}
&&\langle \langle 0|\bar{\psi}\psi |0\rangle _{\mathcal{A-}free}\rangle
_{\varphi }^{LF\ rect}  \notag \\
&=&mN\int_{\delta }^{\Lambda ^{+}}\frac{dp^{+}}{p^{+}}\int_{|p_{\perp }|}%
\frac{d^{2}p_{\perp }}{\left( 2\pi \right) ^{3}}\left[ 1-J_{0}\left( \frac{4g%
}{\varpi }\sqrt{\frac{1}{2N}}\frac{p^{+}\sqrt{p_{\perp }Cp_{\perp }}}{%
p_{\perp }^{2}+m^{2}}\right) \right] .  \label{condensateLFrect}
\end{eqnarray}%
This makes it completely clear how UV sensitivity occurs through large $%
p_{\perp }$ and large $p^{+}$ (and also IR sensitivity through $\delta
\rightarrow 0$).

For a one-loop vacuum-type integral like ours, a commonly used
invariant-mass restriction is to bound $\frac{p_{\perp }^{2}+m^{2}}{p^{+}}%
\leq \Lambda _{inv}$ or, in terms of the on-shell invariant $%
p^{+}p_{on}^{-}=p_{\perp }^{2}+m^{2}$,
\begin{equation}
p_{\perp }^{2}+m^{2}\leq \Lambda _{inv}^{2}.  \label{covariantcutoff}
\end{equation}%
Either way, the integration region is no longer a rectangle but a
\textquotedblleft covariant\textquotedblright\ domain in LF variables. A
clear, explicit choice is%
\begin{equation}
\delta \leq p^{+}\leq \Lambda ^{+},\ p_{\perp }^{2}+m^{2}\leq \Lambda
_{inv}^{2},  \label{covariantdomain}
\end{equation}%
giving%
\begin{eqnarray}
\langle \langle 0|\bar{\psi}\psi |0\rangle _{\mathcal{A-}free}\rangle
_{\varphi }^{LF\;inv} &=&mN\int_{\delta }^{\Lambda ^{+}}\frac{dp^{+}}{p^{+}}%
\int_{p_{\perp }^{2}\leq \Lambda _{inv}^{2}-m^{2}}\frac{d^{2}p_{\perp }}{%
\left( 2\pi \right) ^{3}}  \notag \\
&&\times \left[ 1-J_{0}\left( \frac{4g}{\varpi }\sqrt{\frac{1}{2N}}\frac{%
p^{+}\sqrt{p_{\perp }Cp_{\perp }}}{p_{\perp }^{2}+m^{2}}\right) \right] .
\label{condensatecovar}
\end{eqnarray}%
Let's consider the case of large $\Lambda $ and large $\Lambda ^{+}$ for the
light front cutoff invariant mass. If the transverse polarization/color
structure is isotropic, we can take $C_{ij}=\tau \delta _{ij}$, so $\sqrt{%
p_{\perp }Cp_{\perp }}=\sqrt{\tau }|p_{\perp }|$. We define the momentum
fraction with respect to the external positive scale:%
\begin{equation}
x\equiv \frac{p^{+}}{\varpi }\in \left( 0,1\right) ,\;\frac{dp^{+}}{p^{+}}=%
\frac{dx}{x}.  \label{defx}
\end{equation}%
Then the Bessel argument simplifies nicely and becomes%
\begin{equation}
\left( 4g\sqrt{\tau }\sqrt{\frac{1}{2N}}\right) \frac{x\rho }{\rho ^{2}+m^{2}%
},\;\rho \equiv |p_{\perp }|.  \label{Besselarg}
\end{equation}%
The invariant-mass cutoff for two-particle kinematics is $\frac{m^{2}+\rho
^{2}}{x\left( 1-x\right) }\leq \Lambda ^{2}$, therefore $\;0\leq \rho \leq
\rho _{\max }\left( x\right) \equiv \sqrt{\Lambda ^{2}x\left( 1-x\right)
-m^{2}}$. $\;$This requires $\Lambda ^{2}x\left( 1-x\right) \geq m^{2}$, i.e.%
\begin{equation}
x\in \left[ x_{-},x_{+}\right] ,\;x_{\pm }=\frac{1}{2}\left( 1\pm \sqrt{1-%
\frac{4m^{2}}{\Lambda ^{2}}}\right) ,  \label{xxplusminus}
\end{equation}%
So we also see right away that we need $\Lambda >2m$ for a nonempty domain.
Putting it all together yields the fully explicit invariant-mass--cutoff
result:%
\begin{eqnarray}
&&\langle \langle 0|\bar{\psi}\psi |0\rangle _{\mathcal{A-}free}\rangle
_{\varphi }^{inv.\;mass}  \notag \\
&=&\frac{mN}{4\pi ^{2}}\int_{x_{-}}^{x_{+}}\frac{dx}{x}\int_{0}^{\sqrt{%
\Lambda ^{2}x\left( 1-x\right) -m^{2}}}\rho d\rho \left[ 1-J_{0}\left( \zeta
\frac{x\rho }{\rho ^{2}+m^{2}}\right) \right] ,  \label{condeninvmass}
\end{eqnarray}%
where $\zeta =4g\sqrt{\tau }\sqrt{\frac{1}{2N}}$. The UV region is $\rho \gg
m$. There,%
\begin{equation}
z\left( x,\rho \right) \equiv \zeta \frac{x\rho }{\rho ^{2}+m^{2}}\simeq
\zeta \frac{x}{\rho }\ll 1,  \label{zxro}
\end{equation}%
and the only thing we need is the small-argument expansion $J_{0}\left(
z\right) =1-\frac{z^{2}}{4}+\mathcal{O}\left( z^{4}\right) $, therefore $%
1-\;J_{0}\left( z\right) =\frac{z^{2}}{4}+\mathcal{O}\left( z^{4}\right) $,
and the UV tail of the $\rho $-integral behaves as%
\begin{equation}
\rho d\rho \left[ 1-J_{0}\left( z\right) \right] \simeq \frac{\zeta ^{2}x^{2}%
}{4}\frac{d\rho }{\rho },  \label{rointegral}
\end{equation}%
producing a logarithm cut off at $\rho _{\max }\left( x\right) $. We match
the lower end at $\rho \sim m$ to clearly isolate the UV log (any $\mathcal{O%
}\left( m\right) $ choice shifts only a finite constant). Then, the
UV-controlled part is%
\begin{equation}
\langle \langle 0|\bar{\psi}\psi |0\rangle _{\mathcal{A-}free}\rangle
_{\varphi }^{inv}|_{UV}\simeq \frac{mN}{2\pi ^{2}}\int_{x_{-}}^{x_{+}}\frac{%
dx}{x}\frac{\zeta ^{2}x^{2}}{4}\ln \frac{\rho _{\max }\left( x\right) }{m}.
\label{condeninvUV}
\end{equation}%
We denote $L\equiv \Lambda /m$. Since $\rho _{\max }\left( x\right) =\sqrt{%
L^{2}x\left( 1-x\right) -m^{2}}$,%
\begin{equation}
\ln \frac{\rho _{\max }\left( x\right) }{m}=\frac{1}{2}\ln \left(
L^{2}x\left( 1-x\right) -m^{2}\right) .  \label{lnromax}
\end{equation}%
The subtle NLO effect (next-to-leading order (NLO)) arises because the
invariant-mass cutoff excludes the endpoint regions $x\rightarrow 0,1$: x$%
\in \left[ x_{-},x_{+}\right] $ with $x_{-}\sim 1/L^{2}$. Keeping that exact
and expanding for $L\gg 1$, one finds%
\begin{equation}
\int_{x_{-}}^{x_{+}}dxx\ln \frac{\rho _{\max }\left( x\right) }{m}=\frac{1}{2%
}\ln \frac{\Lambda }{m}-\frac{1}{2}-\frac{m^{2}}{\Lambda ^{2}}\ln \frac{%
\Lambda }{m}+\mathcal{O}\left( \frac{m^{2}}{\Lambda ^{2}}\right) .
\label{intxdxlnromaxm}
\end{equation}%
Substituting this into the condensate yields:%
\begin{equation}
\langle \langle 0|\colon \bar{\psi}\psi \colon |0\rangle _{\mathcal{A}%
}\rangle _{\varphi }^{inv}=\frac{mg^{2}\kappa }{4\pi ^{2}}\left[ \ln \frac{%
\Lambda }{m}-1-2\frac{m^{2}}{\Lambda ^{2}}\ln \frac{\Lambda }{m}+\mathcal{O}%
\left( \frac{m^{2}}{\Lambda ^{2}}\right) \right] +\left(
finite\;IR\;piece\right) .  \label{condenexterinv}
\end{equation}

Let's now examine the hard UV cutoff. When $C_{ij}=\tau \delta _{ij}$, $%
\kappa ^{\mu }=\left( \frac{\varpi }{2},0,0,\frac{\varpi }{2}\right) $ $%
\Rightarrow $ $\kappa ^{+}=\varpi $, $\kappa ^{2}=0$, and transverse
polarization, $\sqrt{\left( \varepsilon ^{a}\cdot p\right) \left(
\varepsilon _{a}\cdot p\right) }=\sqrt{\tau }|p_{\perp }|$, we can rewrite (%
\ref{fercondcos}) as%
\begin{equation}
\langle \langle 0|\bar{\psi}\psi |0\rangle _{\mathcal{A-}free}\rangle
_{\varphi }=2imN\int \frac{d^{4}p}{\left( 2\pi \right) ^{4}}\frac{%
1-J_{0}\left( a\left( p\right) \right) }{p^{2}-m^{2}+i\epsilon },
\label{condensUVcutoff}
\end{equation}%
where%
\begin{equation}
a\left( p\right) =\frac{2g\sqrt{\tau }}{p\kappa }\sqrt{\frac{1}{2N}}%
|p_{\perp }|.  \label{a(p)}
\end{equation}%
Let's point out that%
\begin{equation}
p^{2}-m^{2}+i\epsilon =\left( p^{0}\right) ^{2}-\mathbf{p}%
^{2}-m^{2}+i\epsilon =\left( p^{0}\right) ^{2}-E_{\mathbf{p}}^{2}+i\epsilon
,\;E_{\mathbf{p}}\equiv \sqrt{\mathbf{p}^{2}+m^{2}}.  \label{Ep}
\end{equation}%
The standard identity holds as long as we consider the background factor $%
J_{0}\left( a\left( p\right) \right) $ to be evaluated at the pole (i.e., $%
p^{0}\rightarrow E_{\mathbf{p}}$ in the vacuum contraction):z%
\begin{equation}
\int \frac{dp^{0}}{2\pi }\frac{i}{\left( p^{0}\right) ^{2}-E_{\mathbf{p}%
}^{2}+i\epsilon }=\frac{1}{2E_{\mathbf{p}}}.  \label{integralpzro}
\end{equation}%
So we get%
\begin{equation}
\langle \langle 0|\bar{\psi}\psi |0\rangle _{\mathcal{A-}free}\rangle
_{\varphi }=mN\int \frac{d^{3}p}{\left( 2\pi \right) ^{3}}\frac{1}{E_{%
\mathbf{p}}}\left[ 1-J_{0}\left( a_{on}\left( \mathbf{p}\right) \right) %
\right] ,  \label{condensUVcutoffd3p}
\end{equation}%
where $a_{on}\left( \mathbf{p}\right) $ is $a_{on}\left( p\right) $
evaluated at the on-shell energy $p^{0}=E_{\mathbf{p}}$. With $k^{\mu
}=\left( \frac{\varpi }{2},0,0,\frac{\varpi }{2}\right) $, $p\cdot k=\frac{%
\varpi }{2}\left( p^{0}-p^{3}\right) $. Hence,%
\begin{equation}
a_{on}\left( \mathbf{p}\right) =\beta \frac{|p_{\perp }|}{E_{\mathbf{p}%
}-p^{3}}.  \label{aonp}
\end{equation}%
where $\beta =\frac{4g\sqrt{\tau }}{\varpi }\sqrt{\frac{1}{2N}}\frac{%
|p_{\perp }|}{E_{\mathbf{p}}-p^{3}}$. Note that the would-be collinear
singular point $E_{\mathbf{p}}-p^{3}\rightarrow 0$ only occurs when $%
p_{\perp }\rightarrow 0$ and $m\rightarrow 0$; for $m>0$, $E_{\mathbf{p}%
}-p^{3}=\left( m^{2}+p_{\perp }^{2}\right) /\left( E_{\mathbf{p}%
}+p^{3}\right) $, and the ratio remains finite as $p_{\perp }\rightarrow 0$.
Let $p^{3}\equiv p_{z}$, $|p_{\perp }|\equiv \rho $, so $d^{3}p=2\pi \rho
d\rho dp_{z}$ (in cylindrical coordinates), and the spherical cutoff is $%
\rho ^{2}+p_{z}^{2}<\Lambda ^{2}$. Then%
\begin{eqnarray}
&&\langle \langle 0|\bar{\psi}\psi |0\rangle _{\mathcal{A-}free}\rangle
_{\varphi }^{\left( |\mathbf{p}|<\Lambda \right) }  \notag \\
&=&\frac{mN}{4\pi ^{2}}\int_{-\Lambda }^{\Lambda }dp_{z}\int_{0}^{\sqrt{%
\Lambda ^{2}-p_{z}^{2}}}\frac{\rho d\rho }{\sqrt{\rho ^{2}+p_{z}^{2}+m^{2}}}%
\left[ 1-J_{0}\left( \beta \frac{\rho }{\sqrt{\rho ^{2}+p_{z}^{2}+m^{2}}%
-p_{z}}\right) \right] .  \label{condensateUVcutoff}
\end{eqnarray}

The condensate is influenced by collinear and light-front regions. This is
why light-front cutoffs, invariant-mass regulators, and covariant DR all
give different finite parts. This property makes the object a good way to
test the structure of the light-front vacuum. The condensate shows how a
coherent gluonic background changes the fermion vacuum, just as the
Euler-Heisenberg effective action in QED and fermion polarization in a laser
background do. In this case, though, the background is non-Abelian, color
comes in through $\theta $, and the group structure affects the outcome
through $N$.

\section{Discussion and applications}

In our paper, we used the exact Green function (\ref{Fermionpropmomspace})
for fermions, where the interaction with the external plane-wave background $%
\mathcal{A}_{\mu }^{a}\left( \varphi \right) $ proceeds to all orders by the
dressing matrix $U\left( p,\varphi ,\varphi ^{\prime }\right) $.
Gauge-covariant coupling to the background is guaranteed at all orders since
all interactions with the classical field are represented in $U$. Loop
corrections are still described in powers of $g$, but with background
effects considered exactly instead than as external insertions.

All dressed quantities (propagator, vertex, self-energy, and condensate)
have the same Floquet structure for a periodic plane-wave background. This
appears in the vertex as \ref{fullvertexperiod}. This demonstrates that:
Discrete quanta of momentum $n\kappa $ are exchanged through the background.
Vertex corrections, mass shifts, and condensate oscillations are all
governed by the same harmonic selection rules. In this sense, the background
acts as a coherent medium rather than as a perturbative external insertion.

We have derived the full one-loop, on-shell fermion mass shift in an
external Yang--Mills plane-wave background within the axial gauge, employing
the Mandelstam--Leibbrandt prescription to control spurious gauge
singularities. The final result is \ref{deltamfull}, which encapsulates, in
a gauge-consistent manner, both vacuum renormalization and finite
background-induced effects. An important aspect of the result is the
essential difference between the axial-completion terms $\delta m_{(2+3)}$
and the Feynman-like contribution $\delta m_{(1)}$. While each term depends
on the gauge-defining vectors $n^{\mu }$ and $n^{\ast \mu }$, their sum
yields a well-defined pole mass shift. After cycle averaging over the
plane-wave phases, all terms linear in the background field vanish. The
leading physical correction arises at quadratic order in the background
amplitude and can be written, in the small-field/high-energy limit, as
\begin{equation}
\delta m=\delta m_{free}\left[ 1+\frac{g^{2}}{4N}\frac{\left( \varepsilon
^{a}\cdot p\right) \left( \varepsilon _{a}\cdot p\right) }{\left( p\cdot
\kappa \right) ^{2}}+\mathcal{O}\left( \varepsilon ^{4}\right) \right] .
\label{deltamdeltavac}
\end{equation}%
The correction is suppressed by the invariant $\left( p\cdot \kappa \right)
^{-2}$ and weighted by the color-summed polarization tensor $\varepsilon
_{\mu }^{a}\varepsilon _{\nu a}$, reflecting the non-Abelian nature of the
background. The finite, background-dependent correction shows an interaction
with a classical color field, modifying the fermion's propagation. The
correction is finite and kinematically suppressed.

The explicit axial-gauge formulation and the division into vacuum and
background terms make the result especially suitable for light-front
quantization. The calculation extends the Volkov solution and associated
effective mass concept from QED to non-Abelian gauge theories, offering a
systematic framework for studying fermions in coherent Yang--Mills
backgrounds beyond perturbation theory in the field amplitude. The
background makes finite oscillatory corrections (the phase structure $\theta
\left( p,\varphi \right) $ in $U$). We would like to emphasize that this
feature is crucial because background effects only show up in real,
quantifiable quantities, not in counterterms, and all combined operators
constructed from the dressed propagator are appropriately adjusted. The
exact propagator, effective mass, and vertex together form a closed,
consistent set of building blocks for future studies of quantum processes in
strong, coherent non-Abelian fields. We can use this unified framework to
directly study scattering and radiation processes in coherent Yang--Mills
fields, spin-dependent and polarization-sensitive observables,
effective-field-theory matching in external gauge backgrounds, and
systematic extensions to higher loops or non-equilibrium scenarios.

Our results are relevant to studying heavy-ion collisions, non-Abelian
Schwinger pair production, and early-universe cosmology involving strong
gauge fields \cite{Kharzeev2006}, \cite{Domcke2019}.

\section{Acknowledgement}

This work was partially supported by the grant No. 25RG-1C157 of the Higher
Education and Science Committee of the Ministry of Education, Science,
Culture and Sport RA.

\end{document}